\documentclass[apsrev,prl,amsmath,amssymb,amsfonts,twocolumn,longbibliography]{revtex4-1}
\usepackage{graphicx}
\usepackage{dcolumn}
\usepackage{bm}
\usepackage{amsmath}
\usepackage{amssymb}
\usepackage{mathcomp}
\usepackage{color}
 
\newcommand{\beq}{\begin{equation}}
\newcommand{\eeq}{\end{equation}}
\newcommand{\bea}{\begin{eqnarray}}
\newcommand{\eea}{\end{eqnarray}}

\begin{document}

\title{Page curve entanglement dynamics in an analytically solvable model}

\author{Stefan Kehrein}
\affiliation{Institute for Theoretical Physics, Georg-August-Universit{\"a}t G{\"o}ttingen,
Friedrich-Hund-Platz~1, 37077~G{\"o}ttingen, Germany}

\begin{abstract}
The entanglement entropy of black holes is expected to follow the Page curve. After an initial linear increase with time the entanglement entropy should reach a maximum at the Page time and then decrease. This paper introduces an exactly solvable model of free fermions that explicitly shows such a Page curve: The entanglement entropy vanishes asymptotically for late times instead of saturating at a volume law.
The bending down of the Page curve is accompanied by a breakdown of the semiclassical connection between particle current and entanglement generation, a quantum phase transition in the entanglement Hamiltonian and non-analytic behavior of the $q\rightarrow\infty$ Renyi entropy. These observations are expected to hold for a larger class of systems beyond the exactly solvable model analyzed here.
\end{abstract}

\maketitle

{\it Introduction.} The study of entanglement properties has become an important tool across many fields of physics from condensed matter physics to quantum information theory and black hole physics, bringing about remarkable and fruitful connections between these different fields. One example is the universal area law for ground states of local Hamiltonians \cite{RevModPhys.82.277}, which is e.g.\ central for computational matrix product methods \cite{SCHOLLWOCK201196}, but also for holography \cite{PhysRevLett.96.181602}. Likewise in non-equilibrium much can be learned from the entanglement dynamics: Generically, the entanglement entropy of ergodic systems grows linearly in time until it saturates at a value given by the volume law for excited states. For systems with well-defined quasiparticles the underlying mechanism for this behavior is the local generation of entangled pairs of quasiparticles, whose propagation leads to entanglement of separate spatial regions \cite{Calabrese_2005}. While the general validity of this behavior has been confirmed by many analytical and numerical studies (see e.g.\ Ref.~\cite{PhysRevB.95.094302}), even for systems with diffusive energy transport \cite{PhysRevLett.111.127205}, a long standing debate in black hole physics centers around the very different entanglement dynamics described by the Page curve resulting from the decay of a black hole.  

The first part of the Page curve, that is a linearly increasing entanglement entropy up to the Page time, is perfectly consistent with the above picture: Hawking's semiclassical calculation in~1975 \cite{Hawking_1975} established the production of particle pairs at the event horizon, with one particle escaping to infinity and generating black body radiation with the Hawking temperature. The other particle from the pair falls into the singularity and reduces the mass of the black hole. The entanglement of these particle pairs leads to linearly growing entanglement between the black hole and the enviroment (or equivalently the Hawking radiation). However, as pointed out by Page \cite{PhysRevLett.71.3743,Page_2013}, once the black hole has decayed to about half its original mass, there are not enough degrees of freedom left in the Hilbert space of the black hole to let the entanglement increase even more. In fact, at this so called Page time the entanglement has to decrease again. If one assumes that the black hole was initially formed in a pure state then the entanglement entropy ultimatley has to vanish once the black hole is completely decayed. This behavior is inconsistent with Hawking's semiclassical calculation, and clearly also different from the generic non-equilibrium behavior of a quantum many-body system after a quench discussed above.  

Reproducing the Page curve in a fundamental theory of quantum gravity constitutes one of the major challenges in black hole physics  \cite{Mathur_2009,RevModPhys.93.035002}. The main problem is the observation that the bending down of the Page curve occurs at the Page time where the black hole can still be huge, thereby curvature at the the event horizon being small enough for Hawking's semiclassical calculation in curved space-time to be trustworthy. In other words the bending down of the Page curve occurs at a scale where one expects to understand the relevant laws of nature at the event horizon. This fundamental conflict between the expected bending down of the Page curve and the semiclassical calculation underlies the Hawking information paradox \cite{Mathur_2009,RevModPhys.93.035002} and there have been many attempts to resolve it: Non-unitarity \cite{PhysRevD.14.2460}, small corrections \cite{Papadodimas_2013}, fuzzballs \cite{LUNIN2002342}, firewalls \cite{Almheiri_2013}, most recently the island formula \cite{Almheiri_2019,Pennington_2022} and non-isometric codes \cite{Akers_2022}. 

Motivated by this state of affairs this paper introduces a simple exactly solvable free fermion system plus environment model that shows Page curve like entanglement dynamics. This is interesting from the condensed matter point of view because -- as explained above -- such entanglement behavior is different from the generic picture 
\cite{Calabrese_2005,PhysRevLett.111.127205,PhysRevB.95.094302}. In fact, the dynamics
turns out to be quite intriguing after the equivalent of the Page time defined as the maximum of the entanglement entropy curve.
This exemplifies the difficulty of using non-rigorous arguments for a quantity as subtle as the entanglement entropy. Interestingly, the Page curve dynamics of our model is tied to a quantum phase transition of the entanglement Hamiltonian and non-analytic behavior of the $q\rightarrow\infty$ Renyi entropy at a critical time preceding the Page time.  
In addition, the asymptotic quantum state in the environment has properties which cannot be generated by conventional quench protocols. With respect to black hole physics this paper makes no claim to contribute to the fundamental issue of the Hawking information paradox, although there are some intriguing parallels as will be discussed below. 

System plus bath quantum many-body models with random unitaries that generate a Page curve have previously been discussed in Refs.~\cite{Liu_2021,Lau_2022}. A somewhat similar model to the one in this paper was used in Ref.~\cite{Chen_2020}, however, it yields discontinuous behavior at the Page time and not an actual bending down of the curve. Recent holographic calculations using the island formula have established that this framework describes Page curve physics, though the entanglement entropy becomes constant instead of bending down at the Page time \cite{Chen_2020b,Geng_2021,Geng_2022}. The phenomenon of quantum distillation introduced in Ref.~\cite{Fabian_2009} leads to a decreasing entanglement entropy, but for different reasons than Page curve physics, which is the focus of this paper.

\begin{figure}[t] 
\includegraphics[width=0.95\linewidth]{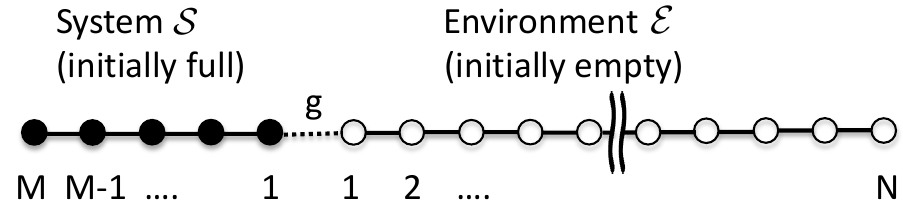}
\caption{\label{fig_reservoir_environment}
System~$\cal S$ coupled to environment $\cal E$. The emptying of the system into the environment leads to Page curve entanglement dynamics. 
}
\end{figure}

{\it Model.}
We consider a free fermion chain, where the first $M$ lattice sites are thought of as a system~$\cal S$
\beq
H_{\rm sys}=t_{\rm sys} \sum_{i=1}^{M-1} \left( c^\dagger_i c_{i+1}+{\rm h.c.}\right)
\eeq
which is weakly coupled to a large environment $\cal E$ with $N\gg M$ sites
\beq
H_{\rm env}=t_{\rm env} \sum_{i=1}^{N-1} \left( f^\dagger_i f_{i+1}+{\rm h.c.}\right)
\label{eq_def_Henv}
\eeq
via 
\beq
H_{\rm c}=g \left( c^\dagger_1 f_1+f^\dagger_1 c_1 \right) 
\label{eq_def_g}
\eeq
with a coupling constant~$g$.
The full Hamiltonian is 
\beq
H=H_{\rm sys}+H_{\rm env}+H_{\rm c}
\label{eq_modelH}
\eeq
and in the initial state the system ${\cal S}$ is completely filled and the environment ${\cal E}$ empty (Fig.~\ref{fig_reservoir_environment})
\beq
|\Psi(0)\rangle = \prod_{i=1}^M c^\dagger_i\,|\Omega\rangle \ .
\label{eq_initialstate}
\eeq
Here $|\Omega\rangle$ is the vacuum.

The total Hilbert space can be decomposed into system and environment sites, 
${\cal H}={\cal H}_{\rm sys}\otimes{\cal H}_{\rm env}$, and we will study the entanglement entropy with respect to this decomposition.
The total state of system plus environment remains pure for all times and therefore the von Neumann entanglement entropy is given by
\bea
S^{\rm (vN)}(t) &=&-{\rm Tr}_{{\cal H}_{\rm env}} \Big( \rho_{\rm env}(t) \ln  \rho_{\rm env}(t) \Big) 
\label{eq_defSent} \\
&=&-{\rm Tr}_{{\cal H}_{\rm sys}} \Big( \rho_{\rm sys}(t) \ln  \rho_{\rm sys}(t) \Big)   \nonumber
\eea
with the reduced density operators $\rho_{\rm env}(t)={\rm Tr}_{{\cal H}_{\rm sys}} \,|\Psi(t)\rangle\langle\Psi(t) |$ and
$\rho_{\rm sys}(t)={\rm Tr}_{{\cal H}_{\rm env}} \,|\Psi(t)\rangle\langle\Psi(t) |$.
The calculation in this paper targets $\rho_{\rm sys}(t)$, for which one has the following intuitive argument why Page curve like behavior is expected:  
The initial particle imbalance between system~$\cal S$ and environment will lead to a decay of the number of particles in the system,
\beq
m(t)\stackrel{\rm def}{=}\langle\Psi(t)|\left( \sum_{i=1}^M c^\dagger_i c_i\right) |\Psi(t)\rangle \ .
\eeq
Initially $m(0)=M$ and after a short transient one expects a constant current $\dot{m}(t)=I$, which can be thought of as the production of a particle-hole pairs at the boundary between system and environment: The holes travels into the system~${\cal S}$ and the particles into the environment~${\cal E}$. Semiclassically particle current and entanglement generation are proportional to one another \cite{Calabrese_2005}, so this leads to linear entanglement growth with a proportionality factor set by the particle current~$I$. At late times the particles will be spread out approximately evenly through system~$\cal S$ and environment~$\cal E$ yielding
\beq
\lim_{t\rightarrow\infty} m(t)\propto \frac{M^2}{M+N} 
\label{eq_longtime_M}
\eeq
which vanishes for $N\gg M^2$. Notice that strictly speaking the long time limit in (\ref{eq_longtime_M}) requires additional averaging over a suitable time window in order to suppress fluctuations, but this detail plays no role in the sequel. Since the system~$\cal S$ empties completely for $N\gg M^2$, the quantum state necessarily has product structure at late times
\beq
|\Psi(t)\rangle = |\Omega_{\rm sys}\rangle \otimes |\Phi_{\rm env}(t)\rangle
\eeq
and therefore vanishing entanglement. This shows why Page curve entanglement dynamics should be expected for this model. 

A more complete picture can be deduced from the observation that any given time one can approximately replace ${\cal H}_{\rm sys}$ 
with the Hilbert space ${\cal H}_{\rm sys}^{\rm (eff)}$ for $m(t)$ spinless fermions on $M$~lattice sites with
\beq
{\rm dim}\: {\cal H}_{\rm sys}^{\rm (eff)} = \left( {M \atop m(t)}\right) \ .
\eeq
This approximation leads to
\bea 
S^{\rm (vN)}(t) &\leq & \ln \, {\rm dim}\: {\cal H}_{\rm sys}^{\rm (eff)} 
\label{eq_Sent_Heff} \\
&\approx & m(t)\,\ln\left(\frac{M}{m(t)}\right) + (M-m(t))\,\ln\left(\frac{M}{M-m(t)}\right) \nonumber
\eea
where we have used the Stirling approximation. This trivially reproduces the short and long time limit, and is consistent with the entanglement entropy being largest when the system~$\cal S$ is half full like for the Page curve. 
Notice that the underlying argument 
of replacing ${\cal H}_{\rm sys}$ with ${\cal H}_{\rm sys}^{\rm (eff)}$ is not rigorous since it neglects fluctuations around the expectation number of particles in the system~$m(t)$. However, for our model (\ref{eq_modelH}) we are able to calculate the entanglement entropy exactly and confirm the intuitive picture developed here. 

{\it Numerical solution.}
The exact calculation of the entanglement entropy relies on the formalism developed by Eisler and Peschel for obtaining the reduced density operator for fermionic or bosonic bilinear Hamiltonians \cite{Peschel_2009}. Applied to our model one has
\beq
\rho_{\rm sys}=K\,\exp\left(-\sum_{i,j=1}^M h_{ij} c^\dagger_i c_j \right)
\eeq
with the matrix $h$ determined by the one-particle correlation matrix $C_{ij}=\langle c^\dagger_i\, c_j \rangle$ via
\beq
h=\ln\Big((1-C)/C\Big) \ .
\label{eq_defh}
\eeq
The diagonalization of $\rho_{\rm sys}$ yields the von Neumann and all other Renyi entanglement entropies. This formalism also applies for bilinear non-equilibrium problems where the one-particle correlation matrix becomes time-dependent~\cite{Peschel_2009}
\beq
C_{ij}(t)=\langle\Psi(0)| c^\dagger_i(t)\, c_j(t)|\Psi(0) \rangle \ .
\eeq
We obtain the time evolved operators $c^\dagger_i(t)$ and $c_j(t)$ from the straightforward exact solution of their Heisenberg equations of motion for the quadratic Hamiltonian~$H$ (\ref{eq_modelH}), which can easily be done for $N=O(10^4)$. This has already been utilized extensively in the literature for various other one-dimensional non-equilibrium problems \cite{Eisler_2008,Eisler_2009b,Eisler_2012,Capizzi_2023}. 

In this paper we use parameters $N=10^4, M=25\ldots 75, g=0.35\ldots 0.8, t_{\rm env}=4$ with the overall energy scale set by the hopping in the system $t_{\rm sys}=1$. The main numerical limitations come from particle reflection at the right boundary of the environment chain, which travel back to the system~$\cal S$ and prevent a further decay of~$m(t)$. This makes it impossible to go to much larger values of~$M$ or smaller values of~$g$ for given~$N=10^4$ if one wants to see Page curve physics before finite size effects become noticeable. All values of the hopping matrix elements $t_{\rm env}, t_{\rm sys}$ lead to Page curve physics; our choice of $t_{\rm env}=4t_{\rm sys}$ is determined by comparison with an analytical approximation that becomes exact for $t_{\rm env}\gg t_{\rm sys}$ (see next section) and the desire to keep finite size effects under control (because the particles in the environment travel and therefore reflect back faster for larger values of~$t_{\rm env}$). Finally the case of a homogeneous chain $t_{\rm sys}=g=t_{\rm env}$ is special since it leads to a logarithmic increase of the entanglement entropy as a function of time vs.\ the generic linear increase for an inhomogeneous chain (this is well known from the literature \cite{Eisler_2012}): Since we are interested in the tunneling limit we use $g<t_{\rm sys}$ and stay away from the special homogeneous point.

\begin{figure}[h] 
\includegraphics[width=0.6\linewidth]{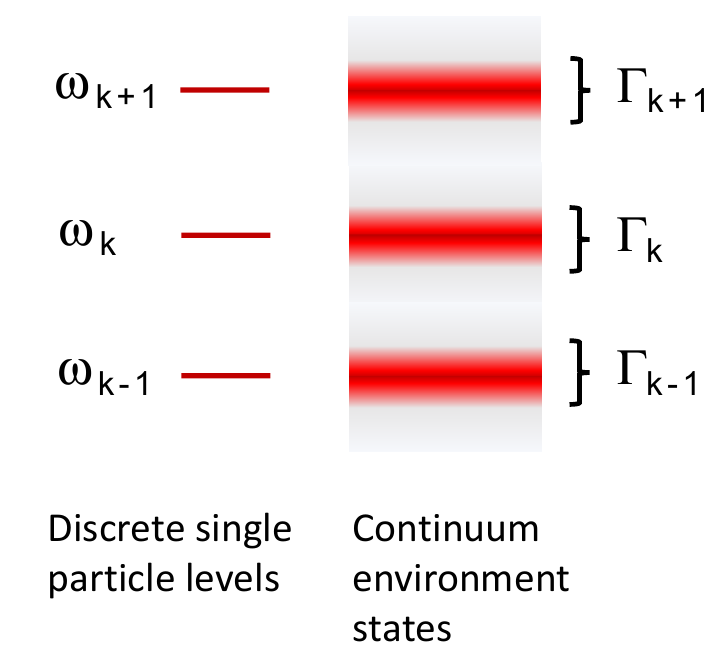}
\caption{\label{fig_RLM}
Equivalent description of the model from Fig.~\ref{fig_reservoir_environment} in terms of single particle levels coupled to a continuum environment (\ref{eq_H_singlelevels}). If condition (\ref{eq_cond_RLM}) is fulfilled the single particle levels couple to effectively disjoint environments, leading to a model of $M$ disjoint resonant level models (\ref{eq_HRLM}).
}
\end{figure}

{\it Analytical solution.} The fact that the Hamiltonian (\ref{eq_modelH}) is quadratic allows for an analytical solution in the weak coupling limit $g\ll t_{\rm sys}, t_{\rm env}$. In a first step one diagonalizes the system Hamiltonian $H_{\rm sys}$, which leads to an equivalent description in terms of single-particle levels 
$c^\dagger_k, c_k$ coupled to the environment,
\beq
H=\sum_{k=1}^M \omega_k \, c^\dagger_k c_k + \sum_{k=1}^M V_k \left( c^\dagger_k f_1+f^\dagger_1 c_k\right) +H_{\rm env} \ .
\label{eq_H_singlelevels}
\eeq
One easily verifies the single particle energies
\beq
\omega_k =-t_{\rm sys}\,\cos\left(\frac{\pi\,k}{M+1}\right)
\label{eq_omegak}
\eeq
and hybridization matrix elements
\beq 
V_k = g\,\sqrt{\frac{2}{M+1}}\,\sin\left(\frac{\pi\,k}{M+1}\right)
\eeq
for $k=1..M$ from the analytical solution of a finite chain. In the limit $N\rightarrow\infty$ one can describe the environment via a continuum density of states $\rho(\epsilon)$. Each single particle level~$\omega_k$ therefore couples predominantly to environment states in an energy interval around $\omega_k$ set by the hybridization
\beq
\Gamma_k(\omega_k)=\pi\rho(\omega_k)\,V_k^2 . 
\label{eq_def_hybridization}
\eeq
If the energy difference between adjacent single particle levels is much larger than the hybridization one can think of these single particle levels as being coupled to disjoint environments, see Fig.~\ref{fig_RLM}. This condition can be satisfied for sufficiently small coupling~$g$ even in the limit $M\rightarrow\infty$ since
\beq
|\omega_{k+1}-\omega_{k}| \propto  \frac{t_{\rm sys}}{M} \gg \Gamma_k(\omega_k) \propto  \frac{g^2}{M} \rho(\omega_k) 
\label{eq_cond_RLM}
\eeq
is true in the weak coupling limit 
\beq
g \ll \sqrt{t_{\rm sys} t_{\rm env}}\ . 
\label{eq_weaklimit}
\eeq
Strictly speaking there is a fraction $g^2/t_{\rm sys}t_{\rm env}$ of single particle levels close to $k=1$ and $k=M$ where condition (\ref{eq_cond_RLM}) does not hold since the dispersion relation (\ref{eq_omegak}) becomes flat, but this is negligible in the limit (\ref{eq_weaklimit}). 

Putting everything together, in the weak coupling limit (\ref{eq_weaklimit}) our model (\ref{eq_H_singlelevels}) can equivalently be written as a collection of $M$~disjoint resonant level models (RLMs)
\beq
H^{\rm (RLM)}_k = \omega_k  c^\dagger_k c_k + \sum_i V_k \left( c^\dagger_k a_{k,i}+a^\dagger_{k,i} c_k\right) +\sum_i \epsilon_i a^\dagger_{k,i} a_{k,i} 
\label{eq_HRLM}
\eeq 
with hybridization (\ref{eq_def_hybridization})
\beq
\Gamma_k(\epsilon) = \frac{2\pi\rho(\epsilon)\,g^2}{M+1}\,\sin^2\left(\frac{\pi\,k}{M+1}\right) 
\eeq
where the operators $a^\dagger_{k,i}, a_{k,i}$ are eigenoperators of the environment Hamiltonian (\ref{eq_def_Henv}).
The reduced density operator of the system $\rho_{\rm sys}(t)$ is then the direct product of reduced density operators of the $M$~resonant level models
\beq
\rho_{\rm sys}(t) = \bigotimes_{k=1}^M \rho^{(RLM)}_k(t)
\eeq
with 
\beq
\rho^{(RLM)}_k(t)=\left( \begin{array}{cc} n_k(t) & 0 \\ 0 & 1-n_k(t) \end{array} \right) 
\label{eq_rhoRLM}
\eeq
where $n_k(t)=\langle\Psi(t)|c^\dagger_k c_k|\Psi(t)\rangle$ is the time-dependent occupation number of the impurity orbital. Initially all impurity orbitals are occupied $n_k(t=0)=1$ and all bath sites are empty. As a function of time the number of particles in the system is given by $m(t)=\sum_{k=1}^M n_k(t)$ and the entanglement entropy
\beq
S^{\rm (vN)}(t)=-\sum_{k=1}^M \left( n_k(t)\,\ln n_k(t) + (1-n_k(t))\,\ln (1-n_k(t)) \right) \ .
\eeq  
In the wide flat band limit $t_{\rm env}\gg t_{\rm sys}$ we can take the bath density of states as being constant $\rho(\omega_k)=\rho$ and it is known that the impurity orbital occupation decays purely exponentially in this wide flat band limit \cite{Guinea_1985,Langreth_1991}
\beq
n_k(t)=e^{-2\Gamma_k t} \ .
\label{eq_expdecay}
\eeq
For large systems $M\gg 1$ one arrives at
\bea
\frac{m(\tau)}{M} &=& \frac{1}{\pi} \int_0^\pi dk\: n_k(\tau) \label{eq_m_tau} 
\label{eq_RLM_m} \\
\frac{S^{\rm (vN)}(\tau)}{M} &=& -\frac{1}{\pi} \int_0^\pi dk\: \Big( n_k(\tau)\,\ln n_k(\tau) \label{eq_Sent_tau} 
\label{eq_RLM_Sent} \\
&&\qquad\qquad + (1-n_k(\tau))\,\ln (1-n_k(\tau)) \Big) \nonumber
\eea
where 
\beq
n_k(\tau)=e^{-\tau\sin^2 k} 
\label{eq_n_tau}
\eeq
with the dimensionless parameter 
\beq
\tau=4\pi\rho g^2 t/M \ .
\label{eq_deftau}
\eeq
Eqs. (\ref{eq_m_tau}--\ref{eq_n_tau}) allow us to plot the entanglement entropy as a function of the fractional decay of the system, 
$S^{\rm (vN)}=S^{\rm (vN)}(m/M)$ such that the parameters $g$ and $\rho$ drop out. The resulting curves (shown in the next section) display the universal behavior in the limit $t_{\rm env}\gg t_{\rm sys}\gg g$, $N,M\rightarrow\infty$ while keeping $N\gg M^2$.

\begin{figure}[h] 
\includegraphics[width=\linewidth]{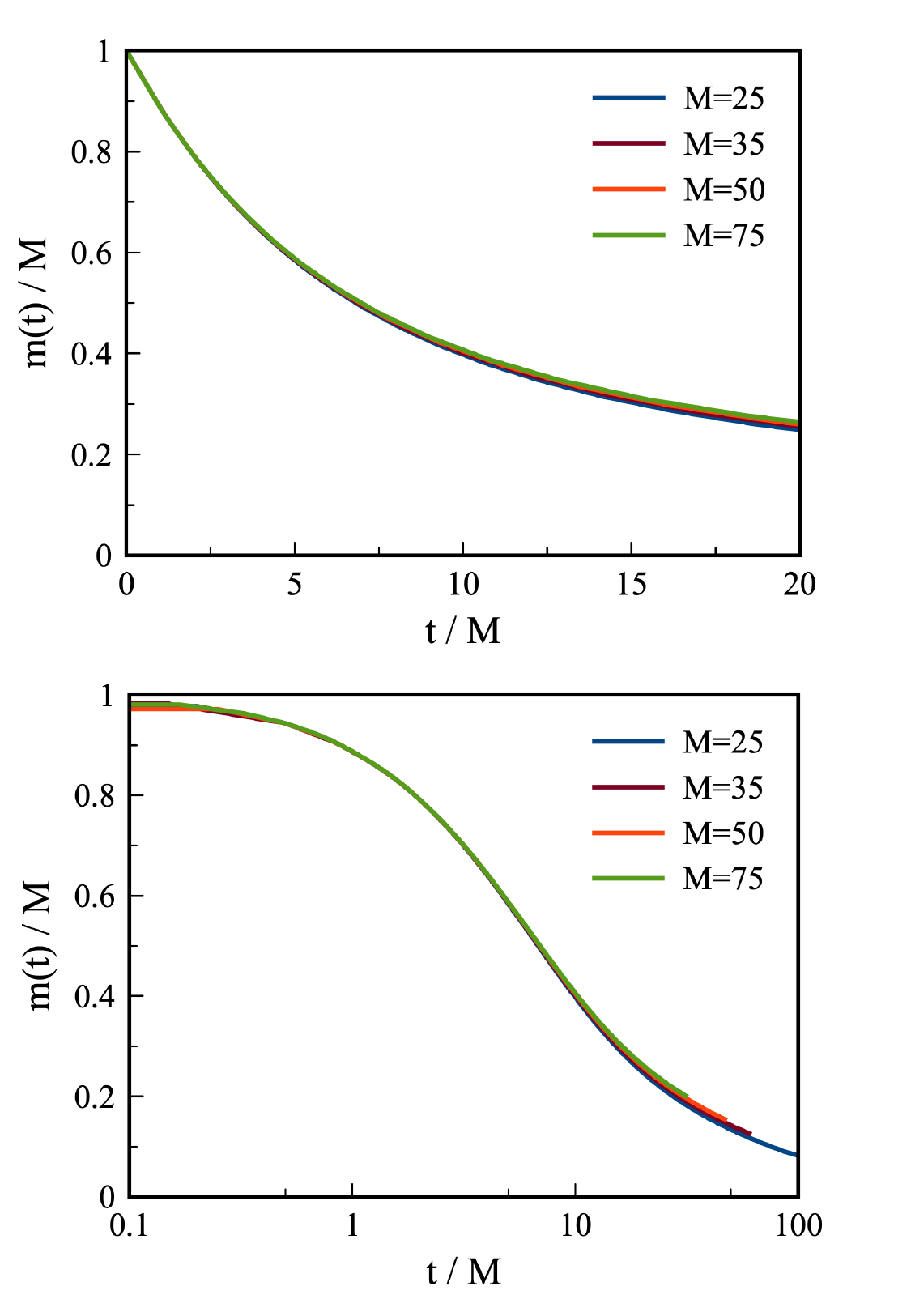}
\caption{\label{fig_M_time}
Decay of the number of particles $m(t)$ in the system~$\cal S$ as a function of time (top: linear scale, bottom: logarithmic scale). The universal behavior after rescaling with $M$ implies a finite current $I(t)=\dot m(t)$ across the boundary for large systems $M\rightarrow\infty$ (always keeping the environment even larger $N\gg M^2$). Parameters: $t_{\rm env}=4, t_{\rm sys}=1, g=0.5, N=10^4$. 
}
\end{figure}

\begin{figure}[h] 
\includegraphics[width=\linewidth]{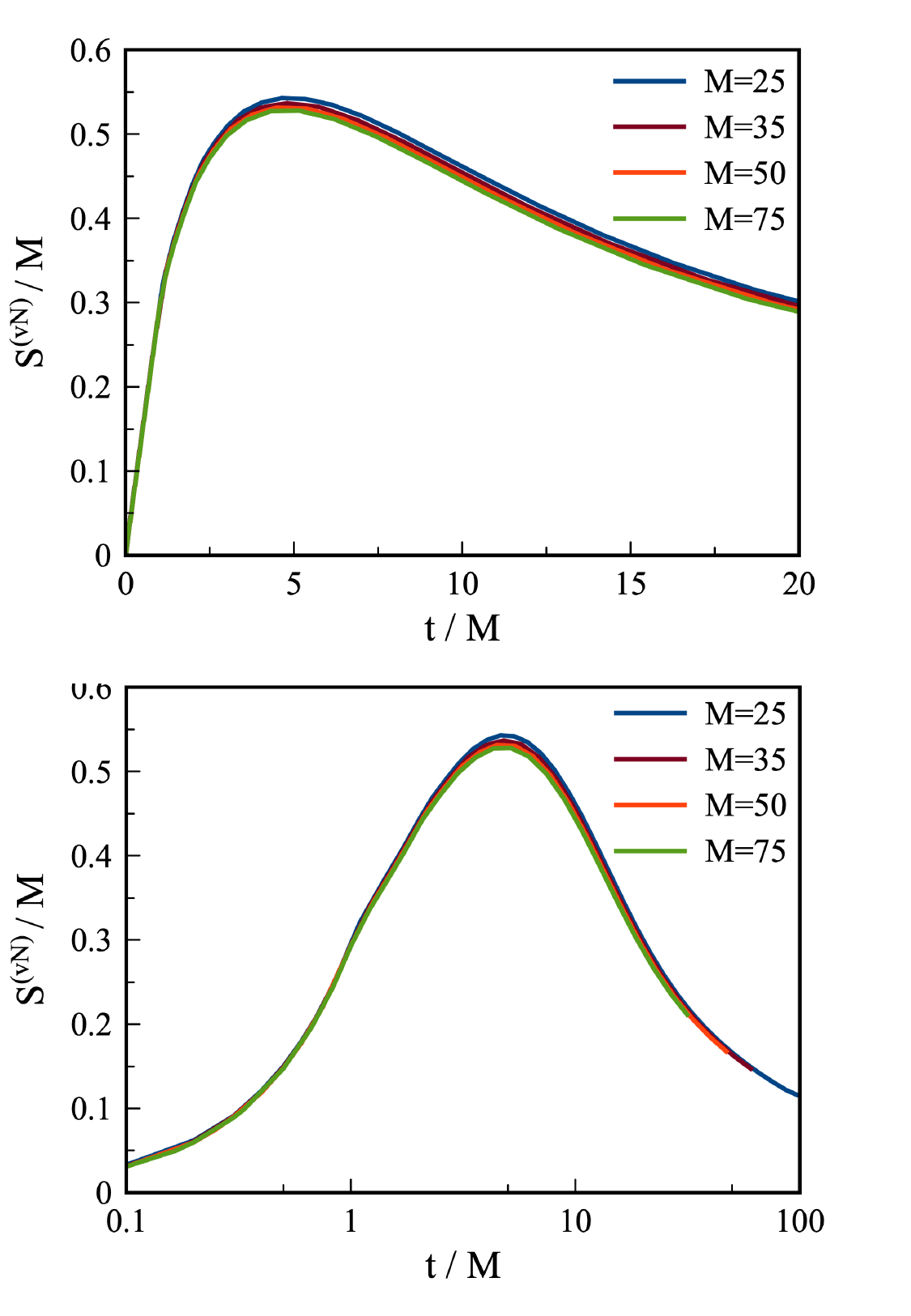}
\caption{\label{fig_SvN_time}
Von Neumann entanglement entropy as a function of time (top: linear scale, bottom: logarithmic scale). The Page curve behavior is clearly visible with the Page time $t_{\rm P}\propto M$ and 
$S^{\rm (vN)}(t_{\rm P})\propto M$.
Parameters: $t_{\rm env}=4, t_{\rm sys}=1, g=0.5, N=10^4$. 
}
\end{figure}

{\it Entanglement entropy.} Starting from the initial state (\ref{eq_initialstate}) the system~$\cal S$ starts to empty as can be seen in Fig.~\ref{fig_M_time} obtained from the numerical solution of the Heisenberg equations of motion. Based on the discussion of (\ref{eq_Sent_Heff}) we therefore expect to see Page curve entanglement dynamics. Fig.~\ref{fig_SvN_time} is obtained from the numerical solution following the Eisler-Peschel formalism described above and 
clearly exhibits this behavior: The entanglement entropy increases approximately linearly at early times, reaches its maximum at the Page time $t_{\rm P}\propto M$ and then decays again. The entanglement entropy at the Page time is proportional to the system size, 
$S^{\rm (vN)}(t_{\rm P})/M \approx 0.53\pm 0.02$. The proportionality factor is smaller than the Page value~$\ln 2$ \cite{PhysRevLett.71.3743}, which is to be expected for a non-interacting system. 

Notice that there is no discernible feature at the Page time in observables like the number of particles in the system $m(t)$, Fig.~\ref{fig_M_time}. One cannot deduce anything about the bending down of the entanglement entropy from $m(t)$ or the particle current across the boundary 
$I(t)=\dot{m}(t)$. This also demonstrates the breakdown of the semiclassical connection between particle current and entanglement generation \cite{Calabrese_2005} 
\beq
\frac{dS^{\rm (vN)}(t)}{dt} \propto I(t)
\label{eq_dSdt}
\eeq
beyond the Page time. Due to the highly entangled state that has dynamically developed up to the Page time one can no longer treat particle-hole production at the boundary as leading to a maximally entangled pair that is not entangled with other particles. 

\begin{figure}[h] 
\includegraphics[width=\linewidth]{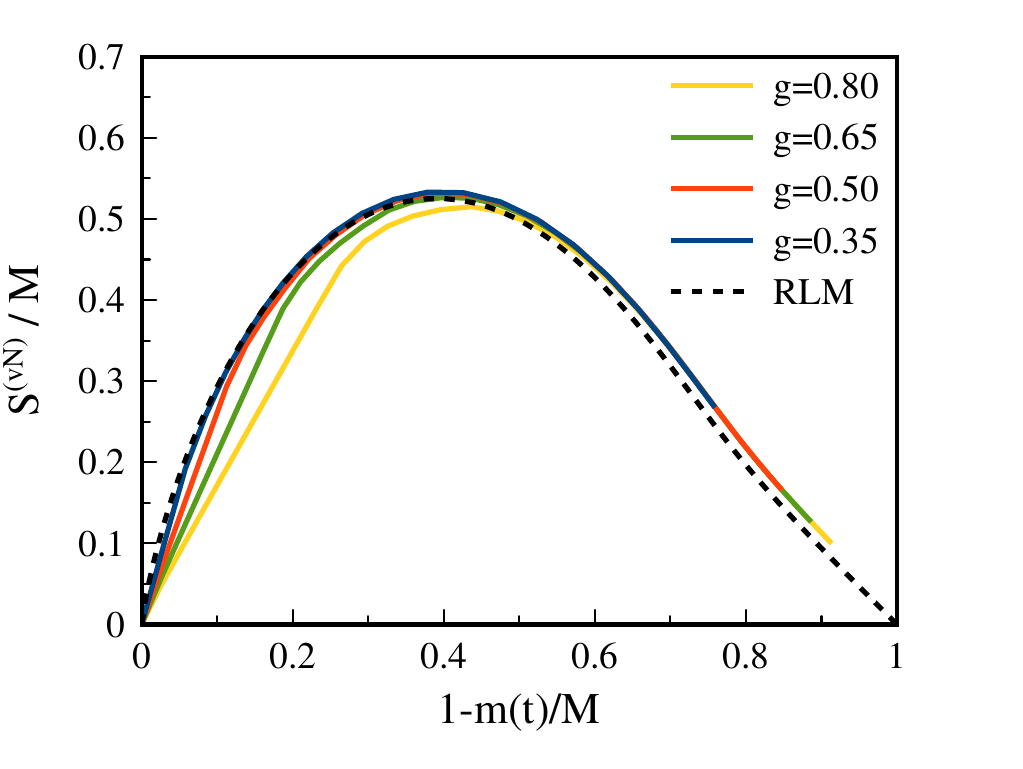}
\caption{\label{fig_SvN_filling}
Von Neumann entanglement entropy as a function of the fraction of particles that have been emitted into the environment for various system-environment couplings~$g$ (\ref{eq_def_g}). The curves become universal in the weak-coupling limit $g\rightarrow 0$ and approach the analytical result from the disjoint resonant level model description (dashed line).
Parameters: $t_{\rm env}=4, t_{\rm sys}=1, M=50, N=10^4$. 
}
\end{figure}

Since the Page time depends on the coupling~$g$ between system and environment, a universal way to depict the Page curve in our model is to plot the entanglement entropy as a function of the fraction of particles emitted into the environment, $1-m(t)/M$. The corresponding curves for different values of~$g$ are shown in Fig.~\ref{fig_SvN_filling}. Specifically, one can observe that the numerical results agree very well with the analytical universal curve from Eqs.~(\ref{eq_m_tau}--\ref{eq_n_tau}) (dashed line) for small values of~$g$. 

The remaining deviations at late times are due to the finiteness of the environment in the numerical simulations: Each eigenmode of the system effectively couples to $N\,\Gamma_k/t_{\rm env}$ modes in the environment, leading to a proportionality factor $(t_{\rm env}/g)^2$ in (\ref{eq_longtime_M}). This implies that the exponential decay (\ref{eq_expdecay}) only persists up to a finite residual occupation scaling like 
$(t_{\rm env}/g)^2\times M/N$, which in turn shifts the finite-$N$ entanglement curves above the RLM result. Going to even larger values of~$N$ decreases these small deviations, but becomes numerically expensive and does not provide insights which seem to justify this effort.
 
\begin{figure}[h] 
\includegraphics[width=\linewidth]{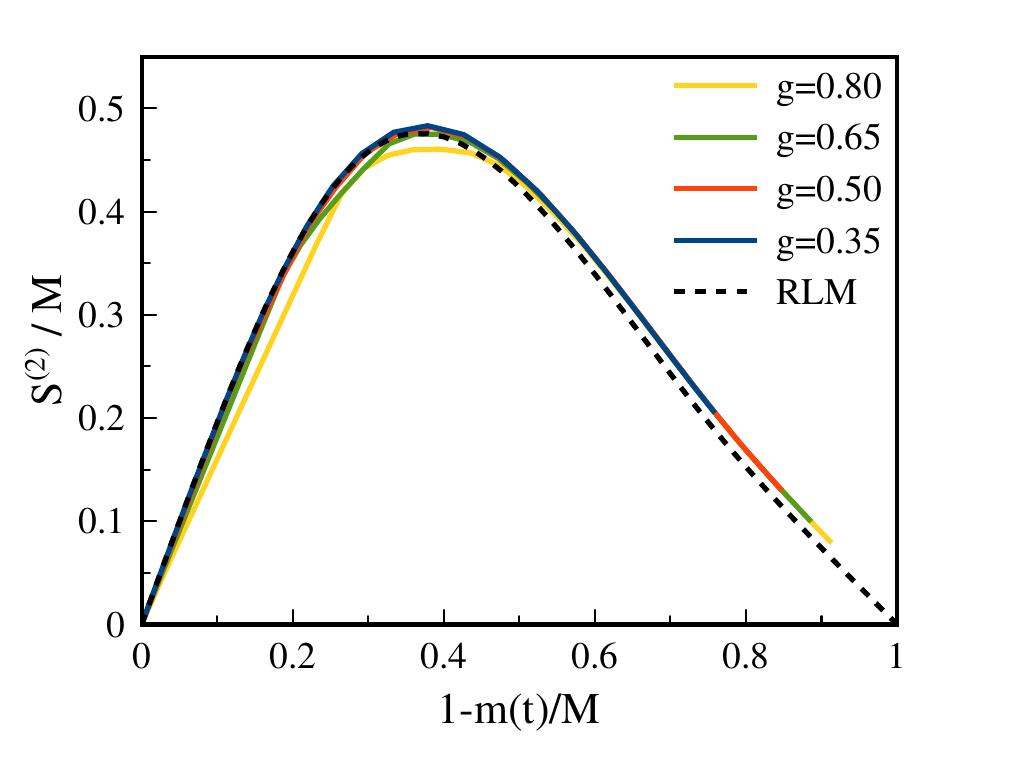}
\caption{\label{fig_Sq2_filling}
Purity $S^{(2)}$ as a function of the fraction of particles that have been emitted into the environment for various system-environment couplings~$g$. The dashed line is the analytical result from the disjoint resonant level model description.
Parameters: $t_{\rm env}=4, t_{\rm sys}=1, M=50, N=10^4$. 
}
\end{figure}

\begin{figure}[h] 
\includegraphics[width=\linewidth]{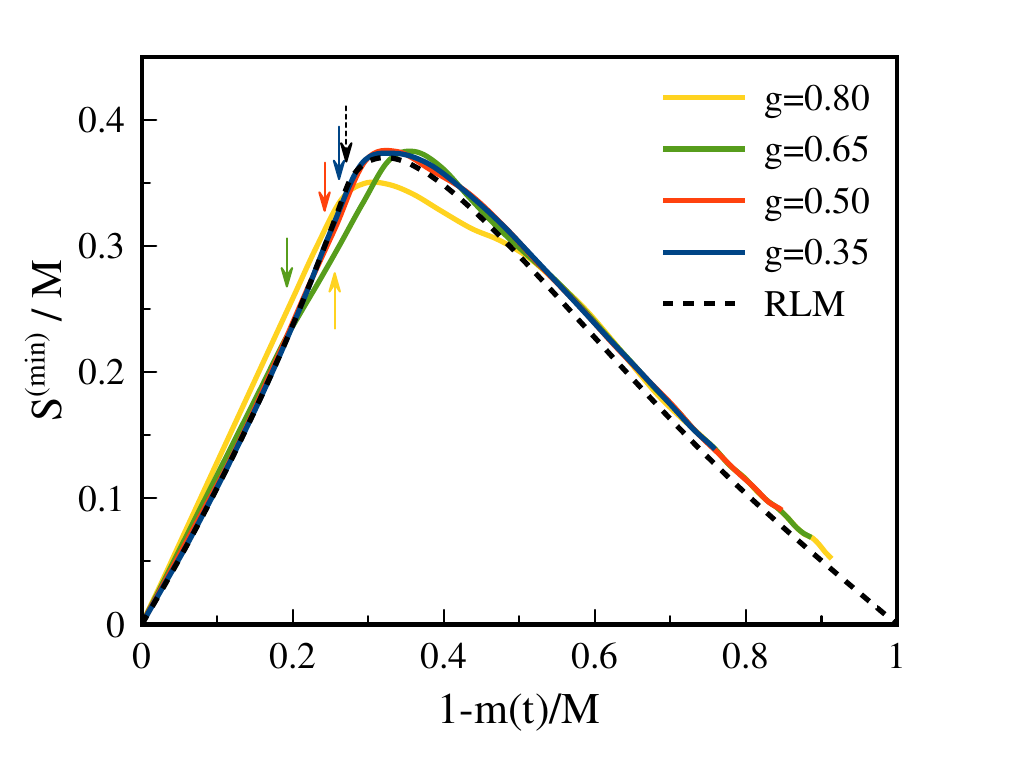}
\caption{\label{fig_Smin_filling}
Min-entropy $S^{({\rm min})}$ as a function of the fraction of particles that have been emitted into the environment for various system-environment couplings~$g$. The dashed line is the analytical result from the disjoint resonant level model description. The arrows mark the critical time with non-analytic behavior of the min-entropy in the resp.~curve.
Parameters: $t_{\rm env}=4, t_{\rm sys}=1, M=50, N=10^4$. 
}
\end{figure}

Page curve behavior is also visible in all Renyi entropies 
\beq
S^{(q)}=\frac{1}{1-q}\,\ln{\rm Tr}_{{\cal H}_{\rm sys}}\left(\rho_{\rm sys}\right)^q \ .
\label{eq_defRenyi}
\eeq
This is depicted in Fig.~\ref{fig_Sq2_filling} for the purity~$S^{(2)}(t)$ and in Fig.~\ref{fig_Smin_filling} for the 
min-entropy~$S^{({\rm min})}(t)=\lim_{q\rightarrow\infty} S^{(q)}(t)$. For $q\rightarrow 1$ one obtains the usual von Neumann entanglement entropy $S^{\rm (vN)}(t)=\lim_{q\rightarrow 1} S^{(q)}(t)$.
Notice that the min-entropy is just determined by the largest eigenvalue of the Schmidt decomposition 
\beq
S^{({\rm min})}(t)=-\ln\lambda_1(t) 
\label{eq_Smin}
\eeq
with
\beq
|\Psi(t)\rangle=\sum_{i=1}^M \sqrt{\lambda_i} \, |\phi_{{\rm sys},i}(t)\rangle \otimes |\Phi_{{\rm env},i}(t)\rangle
\label{eq_Schmidt}
\eeq
and $\lambda_1\geq\lambda_2\geq \ldots \geq \lambda_M$.
Interestingly, the min-entropy becomes non-analytic at the time indicated by the arrows in Fig.~\ref{fig_Smin_filling}. This will be discussed in detail in the next section.

\begin{figure}[h] 
\includegraphics[width=\linewidth]{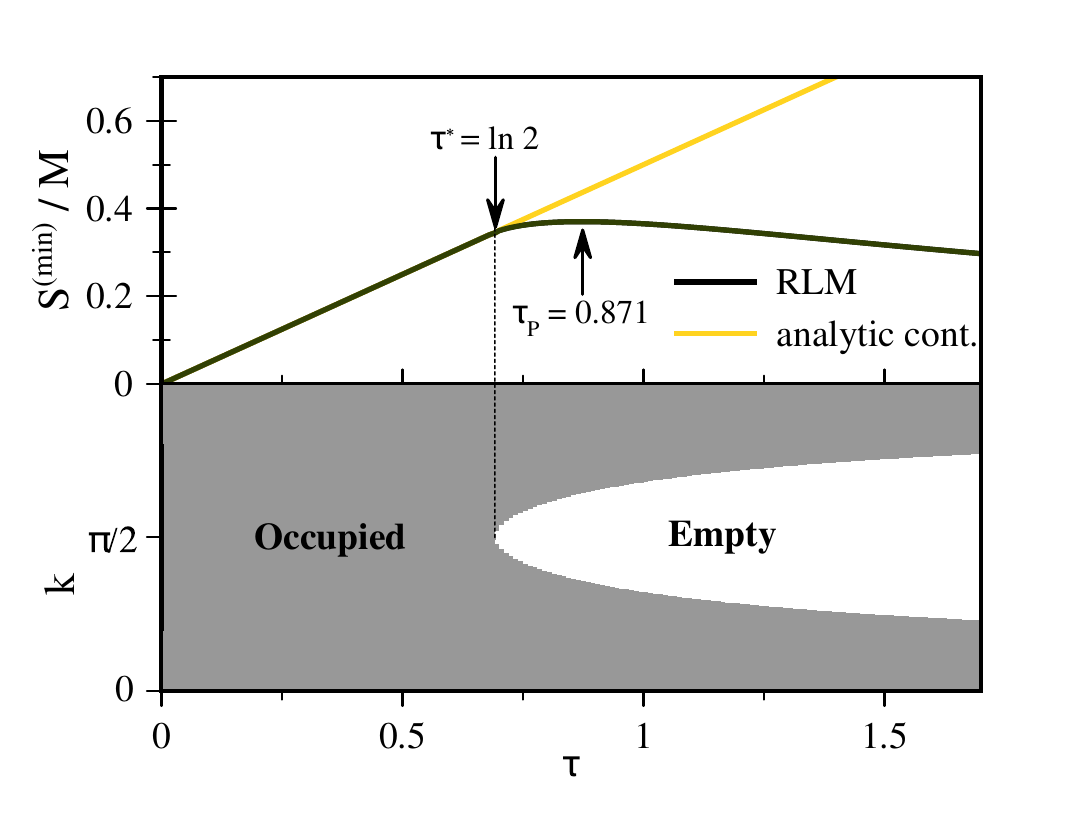}
\caption{\label{fig_Smin_transition}
Upper panel: The black line shows the min-entropy $S^{({\rm min})}$ of the disjoint resonant level model description as a function of dimensionless time~$\tau$ from (\ref{eq_deftau}). The yellow line is the analytic continuation based on the early time behavior according to (\ref{eq_Smin_ana}). Lower panel: The dark region depicts values of the wave vector~$k$ with $n_k(\tau)>1/2$, the light region values of~$k$ with $n_k(\tau)<1/2$.
}
\end{figure}

{\it Non-analytic behavior of $S^{({\rm min})}$.} The non-analytic behavior of $S^{({\rm min})}$ at a critical time (or equivalently: at a critical filling) is difficult to discern from Fig.~\ref{fig_Smin_filling}. However, it can easily be deduced from analytical considerations. Let us first analyze the disjoint resonant level model description. Eqs.~(\ref{eq_rhoRLM}) and (\ref{eq_Smin}) imply for large~$M$
\beq
\frac{S^{(\rm min)}(\tau)}{M}=-\frac{1}{\pi}\int_0^\pi dk\,\ln\Big({\rm max}\left(n_k(\tau),1-n_k(\tau)\right)\Big) \ .
\eeq
This function is plotted in Fig.~\ref{fig_Smin_transition}. For $\tau\leq\tau^*=\ln 2$ one has $n_k(\tau)\geq 1/2$ for all~$k\in [0,\pi]$ and therefore
\beq
\frac{S^{(\rm min)}(\tau)}{M}=\frac{\tau}{2} \ .
\label{eq_Smin_ana}
\eeq
For $\tau>\tau^*$ one has instead $1-n_k(\tau)\geq 1/2$ for the interval~$I_{\rm empty}$ of~$k$-values centered around~$\pi/2$ (lower panel of Fig.~\ref{fig_Smin_transition}) because the decay rates $\Gamma_k$ are largest in the middle of the band,
\beq
I_{\rm empty}=\{k\in [0,\pi] \:|\: n_k(\tau)<1/2\} \ .
\eeq
For $\tau>\tau^*$ one can write 
\beq
\frac{S^{(\rm min)}(\tau)}{M}=\frac{\tau}{2} - k(\tau)
\label{eq_Smin_ktau}
\eeq
with
\bea
k(\tau)&=&-\frac{1}{\pi}\int_{I_{\rm empty}} dk\:\Big( \ln n_k(\tau)-\ln (1-n_k(\tau)) \Big) \\
&=& \tau\,\frac{1}{\pi}\int_{I_{\rm empty}} dk\: \sin^2k \nonumber \\
&&+ \frac{1}{\pi}\int_{I_{\rm empty}} dk\: \ln \left(1-e^{-\tau\,\sin^2k}\right) 
\eea
An expansion around $\tau=\tau^*+\delta\tau=\ln 2+\delta\tau$, $\delta\tau\geq 0$ yields
\beq
k(\tau^*+\delta\tau)=k_0\,\delta\tau^{3/2} +O(\delta\tau^{5/2}) 
\eeq
with $k_0=8/3\pi\sqrt{\ln 2}$.

This implies that 
$S^{(\rm min)}(\tau)$ is non-analytic at $\tau=\tau^*$ as clearly visible in Fig.~\ref{fig_Smin_transition} in the sense that the second 
time derivative becomes discontinuous,
\bea
\lim_{\tau\uparrow\tau^*} \frac{d^2}{d\tau^2}\,\left(\frac{S^{(\rm min)}(\tau)}{M}\right) &=& 0 \nonumber \\
\lim_{\tau\downarrow\tau^*} \frac{d^2}{d\tau^2}\,\left(\frac{S^{(\rm min)}(\tau)}{M}\right) &=& \infty \ .
\label{eq_2ndderiv}
\eea
The non-analytic behavior at~$\tau^*=\ln\, 2$ marks a well-defined dynamical transition and precedes the Page time~$\tau_{\rm P}$ defined as the time where the entanglement entropy reaches its maximum (notice that this definition of the Page time depends on~$q$, so here we are referring to $q=\infty$). From (\ref{eq_RLM_m}) and (\ref{eq_RLM_Sent}) one can derive the corresponding fillings for comparison with Fig.~\ref{fig_Smin_filling}: $\frac{m(\tau^*)}{M}=0.728$ and $\frac{m(\tau_{\rm P})}{M}=0.678$. 

While studies of the entanglement entropy often focus on the von Neumann entanglement entropy, the min-entropy also has an 
important quantum information theoretic interpretation: $S^{({\rm min})}$ determines the one-shot distillable entanglement \cite{Brandao_2011}
and gives the distance to the closest separable state 
\beq
|D(t)\rangle=|\phi_{\rm sys}(t)\rangle \otimes |\Phi_{\rm env}(t)\rangle
\eeq
in the sense of the Fubini-Study metric \cite{Lockhart2002}: If one chooses $|D(t)\rangle$ such that $|\langle D(t)|\Psi(t)\rangle|^2$ is maximized then
\beq
-\ln \, |\langle D(t)|\Psi(t)\rangle|^2 = S^{\rm (min)}(t) \ .
\eeq
In the Schmidt decomposition (\ref{eq_Schmidt}) this means 
\beq
|D(t)\rangle=|\phi_{{\rm sys},1}\rangle \otimes |\Phi_{{\rm env},1}\rangle \ .
\eeq
So an increasing min-entropy of $|\Psi(t)\rangle$ as a function of time means a larger distance to the closest unentangled state 
$|D(t)\rangle$. For $\tau<\tau^*$ the closest separable state is the one with all resonant level modes occupied
\beq
|\phi_{{\rm sys},1}(\tau)\rangle = \prod_{k=0}^\pi c^\dagger_k\,|\Omega_{\rm sys}\rangle \ ,
\eeq
which is actually time independent for $\tau<\tau^*$. For $\tau>\tau^*$ the closest separable state has the system component
\beq
|\phi_{{\rm sys},1}(\tau)\rangle = \prod_{k\in I_{\rm occ}(\tau)} c^\dagger_k\,|\Omega_{\rm sys}\rangle 
\label{eq_closest}
\eeq
with the interval $I_{\rm occ}(\tau)=\{k\in [0,\pi] \:|\: n_k(\tau)\geq 1/2\}$. This means that some of the modes of the closest unentangled state are now
empty as depicted in the lower panel of Fig.~\ref{fig_Smin_transition}. The sudden appearance of a different closest separable state at 
$\tau=\tau^*$ is responsible for the dynamic transition and non-analytic behavior of $S^{\rm (min)}(\tau)$ at $\tau=\tau^*$.

{\it Quantum phase transition of the entanglement Hamiltonian.}
Another point of view of the dynamical transition at~$\tau^*$ can be obtained from the entanglement Hamiltonian (also called the modular Hamiltonian) defined via
\beq
\rho_{\rm sys}(t) = \exp\left(-H_{\rm ent}(t)\right) \ .
\label{eq_defHent}
\eeq
Defining the partition function
\beq
Z_{\rm ent}(\beta)={\rm Tr}_{{\cal H}_{\rm sys}} \exp\left(-\beta\,H_{\rm ent}\right) 
\eeq
and
\beq 
F_{\rm ent}(\beta)=-\beta^{-1}\,\ln\,Z_{\rm ent}(\beta)
\eeq
one can immediately derive from (\ref{eq_defRenyi})
\beq
S^{(q)}=\frac{q}{q-1}\,F_{\rm ent}(q) \ .
\label{eq_relSF}
\eeq
So the Renyi entropies can be thought of as being proportional to the free energy of the entanglement Hamiltonian at inverse temperature 
$\beta=q$. Therefore the min-entropy with $q\rightarrow\infty$ corresponds to the zero temperature limit and is just the ground state energy of the entanglement Hamiltonian
\beq
S^{\rm (min)}(t)=E^{\rm GS}_{\rm ent}(t) \ .
\label{eq_Smin_EGS}
\eeq
Hence non-analytic behavior of the min-entropy is equivalent to non-analytic behavior of the ground state energy of the entanglement Hamiltonian, that is a quantum phase transition. Notice that because of (\ref{eq_defHent}) one has $Z(1)=1$, which implies 
$E^{\rm GS}_{\rm ent}(t)\geq 0$. Also one needs to be careful when taking the limit $q\rightarrow 1$ in (\ref{eq_relSF}) in order to obtain the von Neumann entanglement entropy.

For the disjoint resonant level model it is straightforward to derive the entanglement Hamiltonian from (\ref{eq_rhoRLM}),
\beq
H_{\rm ent}(\tau)=\sum_k \nu_k(\tau)\, c^\dagger_k c_k + a(\tau) \ .
\eeq
Here the single particle energies are given by
\beq
\nu_k(\tau) = \ln\left( e^{\tau\sin^2 k} - 1\right)
\eeq
and the constant
\beq
a(\tau)=-\sum_k \ln\left(1-e^{-\tau\sin^2 k} \right) \ .
\eeq
For $\tau<\tau^*$ one has $\nu_k(\tau)<0$ for all~$k$. Therefore all modes are occupied and the ground state energy is
\beq
E^{\rm GS}_{\rm ent}(\tau) =a(\tau)+\sum_k \nu_k(\tau) \ .
\label{eq_EGS_before}
\eeq
Once $\tau>\tau^*$ there are
unoccupied modes with $\nu_k(\tau)>0$,
\beq
E^{\rm GS}_{\rm ent}(\tau) =a(\tau)+\sum_{k:\nu_k(\tau)<0} \nu_k(\tau) \ .
\label{eq_EGS_after}
\eeq
One can easily verify that (\ref{eq_EGS_before}) agrees with (\ref{eq_Smin_ana}) and that (\ref{eq_EGS_after}) agrees with (\ref{eq_Smin_ktau}) as expected from (\ref{eq_Smin_EGS}), so one has a quantum phase transition of the entanglement Hamiltonian at~$\tau=\tau^*$. This is accompanied by non-analytic behavior of the particle number in the ground state of the entanglement Hamiltonian,
\beq
m^{\rm GS}_{\rm ent}\stackrel{\rm def}{=} \langle {\rm GS}_{\rm ent}\, |\, \sum_k c^\dagger_k \,c_k \, | \, {\rm GS}_{\rm ent}\rangle \ .
\eeq
One has
\beq
m^{\rm GS}_{\rm ent}(\tau) = \sum_{k\in I_{\rm occ}(\tau)} 1
\eeq
yielding 
\beq
\frac{m^{\rm GS}_{\rm ent}(\tau)}{M} = \left\{ \begin{array}{ll} 1 &\quad \tau\leq\tau^* \\ \frac{2}{\pi}\,\arcsin\sqrt{\frac{\tau^*}{\tau}} & \quad \tau>\tau^* 
\end{array} \right.
\label{eq_mGS}
\eeq 
This expression just corresponds to the occupied region in the lower panel of Fig.~\ref{fig_Smin_transition}. At the transition one finds
\beq
\frac{m^{\rm GS}_{\rm ent}(\tau^*+\delta\tau)}{M} =1-\frac{2}{\pi\sqrt{\ln 2}}\,\delta\tau^{1/2} +O(\delta\tau^{3/2}) 
\eeq
for $\delta\tau>0$.

\begin{figure}[h] 
\includegraphics[width=\linewidth]{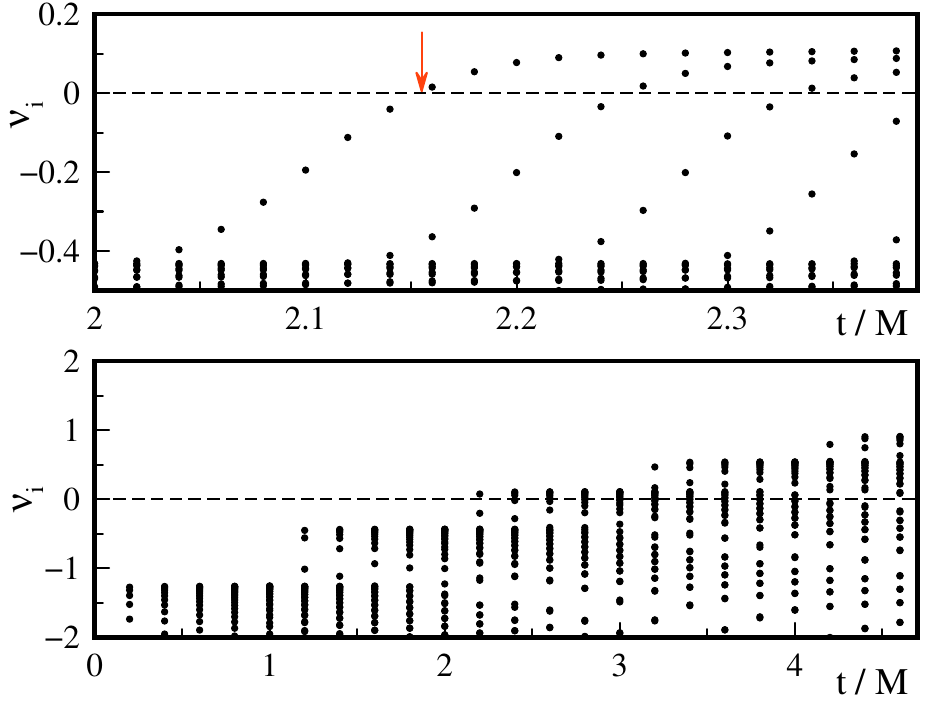}
\caption{\label{fig_EntanglementSpectrum}
Single particle energies $\nu_i$ of the entanglement Hamiltonian as a function of time. The overall trend is that the single particle energies become larger for later times since this corresponds to the emptying of the system into the environment. The upper panel shows a zoomed in plot where one can more clearly see when the first eigenvalue crosses zero (marked by red arrow), corresponding to the quantum phase transition of $H_{\rm ent}(t)$. 
Parameters: $t_{\rm env}=4, t_{\rm sys}=1, g=0.5, M=50, N=10^4$. 
}
\end{figure}

The entanglement Hamiltonian also allows one to identify the critical times in the dynamics away from the small~$g$ limit, see Fig.~\ref{fig_Smin_filling}. Following Ref.~\cite{Peschel_2009} the single-particle energies~$\nu_i$ of the entanglement Hamiltonian
are the eigenvalues of~$h$ from~(\ref{eq_defh}), so the quantum phase transition of $H_{\rm ent}(t)$ is determined by the
first eigenvalue crossing zero since this implies that $m^{\rm GS}_{\rm ent}(\tau)$ drops from~$M$ to~$M-1$. This is depicted in 
Fig.~\ref{fig_EntanglementSpectrum} and this analysis located the critical times/fillings in Fig.~\ref{fig_Smin_filling}.

Looking at the Page curve from the point of view of the entanglement Hamiltonian also leads to a better understanding of the behavior of the finite-$q$ Renyi entropies~$S^{(q)}$. According to (\ref{eq_relSF}) these can be identified with the free energy at inverse temperature~$\beta=q$. Since the quantum phase transition of $H_{\rm ent}(t)$ is due to level crossing between sectors with different particle numbers, non-analytic behavior only occurs at zero temperature, that is in the min-entropy. All the finite-$q$ Renyi entropies remain analytic at all times, with the non-analytic behavior at the critical time emerging in the zero temperature ($q\rightarrow\infty$) limit. 
So while the critical behavior of $S^{\rm (min)}$ does not extend to finite~$q$ \cite{Chandran2014}, the bending down of the finite-$q$ entanglement entropies can continuously be traced back to the quantum phase transition at~$q=\infty$ and the non-analytic behavior of the ground state particle number (\ref{eq_mGS}).

{\it Low energy variance states.} Behavior related to the Page curve can also be seen in the energy variance of the environment Hamiltonian (\ref{eq_def_Henv}),
\beq
\Delta H_{\rm env}^2(t) \stackrel{\rm def}{=} \langle \big( H_{\rm env}-\langle H_{\rm env}\rangle_t \big)^2 \rangle_t \ .
\eeq
Here $\langle\ldots\rangle_t$ indicates that the expectation value is taken with respect to the state at time~$t$,
$\langle O\rangle_t = \langle \Psi(t)|O|\Psi(t)\rangle$.
$\Delta H_{\rm env}(t)$ is indicative of the number of eigenstates that contribute to the state in the environment. It can only vanish if the environment is in an eigenstate of~$H_{\rm env}$. Generically, for independent unentangled particles the variance $\Delta H_{\rm env}^2$ is proportional to their number. Interestingly, this is very different in our model where one generates a low variance state with $\Delta H_{\rm env}^2(t)\propto O(M^0)$ for $t\rightarrow\infty$. The argument for this is similar to the one for the Page curve: The variance of the full Hamiltonian $H=H_{\rm sys}+H_{\rm c}+H_{\rm env}$ is time invariant and therefore given by its initial value, to which only the boundary coupling term contributes,
\bea
\Delta H^2(t) &=& \Delta H^2(t=0) = \langle \big( H_{\rm c}-\langle H_{\rm c}\rangle_0 \big)^2 \rangle_0 \nonumber \\
&=& g^2 \ .
\eea
Some straightforward manipulation yields
\bea
\Delta H_{\rm env}^2(t) 
&=& \Delta H^2(t) \nonumber \\
&&+\Delta H_A^2(t) +2\langle H\rangle_0\,\langle H_A\rangle_t \nonumber \\
&& -\langle\big\{ H,e^{iHt} H_A e^{-iHt} \big\}\rangle_0 \ ,
\label{eq_DeltaHenv}
\eea
where $H_A=H_{\rm sys}+H_{\rm c}$. Since the system~$\cal S$ empties asympotically for large environments $N\gg M^2$, the second and third line of (\ref{eq_DeltaHenv}) vanish for $t\rightarrow\infty$ and hence
\beq
\lim_{t\rightarrow\infty} \Delta H_{\rm env}^2(t) = \Delta H^2(t)  = g^2 \ .
\label{eq_low_variance}
\eeq

\begin{figure}[h] 
\includegraphics[width=\linewidth]{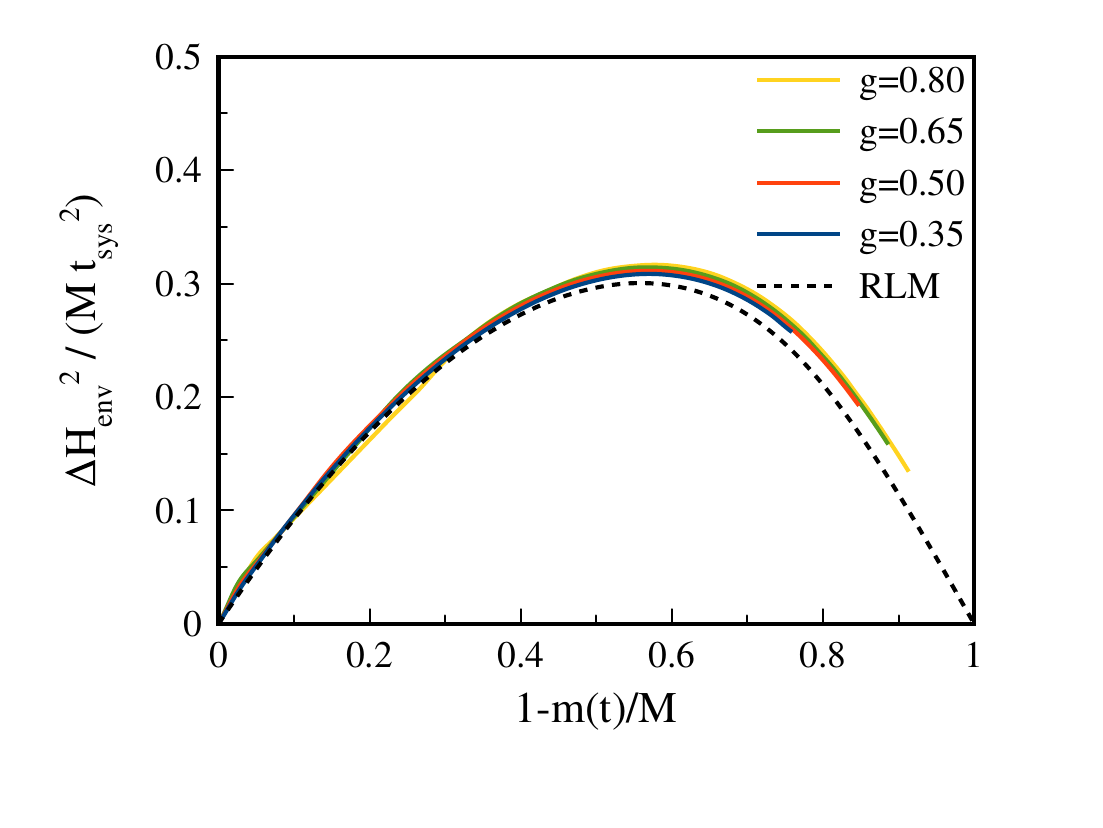}
\caption{\label{fig_DeltaHenv_filling}
Energy variance with respect to the environment Hamiltonian (\ref{eq_def_Henv}) as a function of the fraction of particles that have been emitted into the environment for various system-environment couplings~$g$. The dashed line is the analytical result from the disjoint resonant level model description with a limiting value 
$\lim_{t\rightarrow\infty} \Delta H_{\rm env}^2(t)=O(M^0)$ corresponding to the generation of a low energy variance state in the environment.
Parameters: $t_{\rm env}=4, t_{\rm sys}=1, M=50, N=10^4$. 
}
\end{figure}

Fig.~\ref{fig_DeltaHenv_filling} depicts the scaled energy variance $\Delta H_{\rm env}^2(t)/M$ for various couplings~$g$ from the numerical solution of
(\ref{eq_DeltaHenv}). One observes that the energy variance $\Delta H_{\rm env}^2(t)$ reaches it largest value proportional to system size~$M$ when about half the number of particles has been emitted into the environment, after which the energy variance becomes smaller again. For weak system-environment coupling~$g$ one can also the mapping to the model of disjoint resonant level models (Fig.~\ref{fig_RLM}),
\beq
\Delta H_{\rm env}^2(t) = \sum_{k=1}^M \left( \Delta H_k^{\rm (RLM)}\right)^2(t) 
\eeq
with
\beq
 \left( \Delta H_k^{\rm (RLM)}\right)^2(t) =\omega_k^2\,n_k(t)\,(1-n_k(t)) +O(V_k^2) \ .
 \eeq
For $M\gg 1$ this gives 
\beq
\frac{\Delta H_{\rm env}^2(\tau)}{M} =\frac{t_{\rm sys}^2}{\pi} \int_0^\pi dk\,\cos^2 (k) \, n_k(\tau)\,(1-n_k(\tau)) 
\eeq
with $n_k(\tau)$ from (\ref{eq_n_tau}). The resulting curve is depicted in Fig.~\ref{fig_DeltaHenv_filling}: It agrees very well with the numerical solution and shows the generation of a low variance state in the environment as the system~$\cal S$ continues to emit particles into the environment after the Page time.

\begin{figure}[h] 
\includegraphics[width=\linewidth]{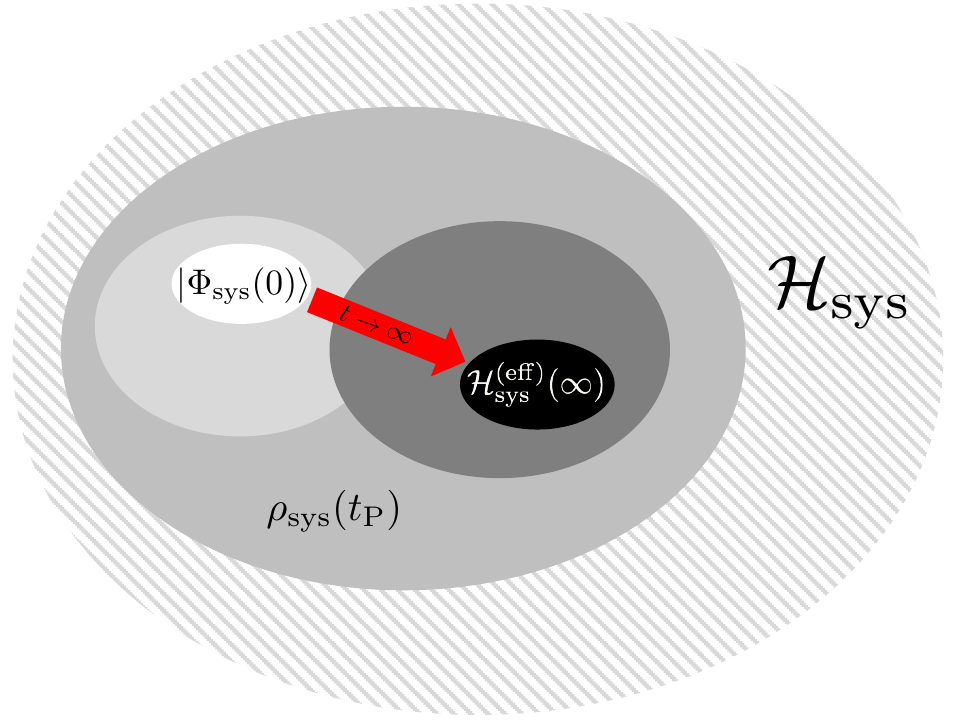}
\caption{\label{fig_effH}
Schematic picture of the dynamics leading to the Page curve. The full colored regions depict the size of the effective Hilbert space  that is necessary in order to represent $\rho_{\rm sys}(t)$ at different times~$t$. Darker colors correspond to later times. In our model eventually the particle current forces the system~$\cal S$ into a small effective Hilbert space ${\cal H}^{\rm (eff)}_{\rm sys}(\infty)$. Although the state $\rho_{\rm env}(t)$ in the environment at late times has a complicated non-local structure, its von Neumann entropy is bounded by $\ln {\rm dim} {\cal H}^{\rm (eff)}_{\rm sys}(\infty)$. Similar scenarios could also be realized in other models by forcing mechanisms like energy or spin currents.
}
\end{figure}

\begin{figure}[h] 
\includegraphics[width=\linewidth]{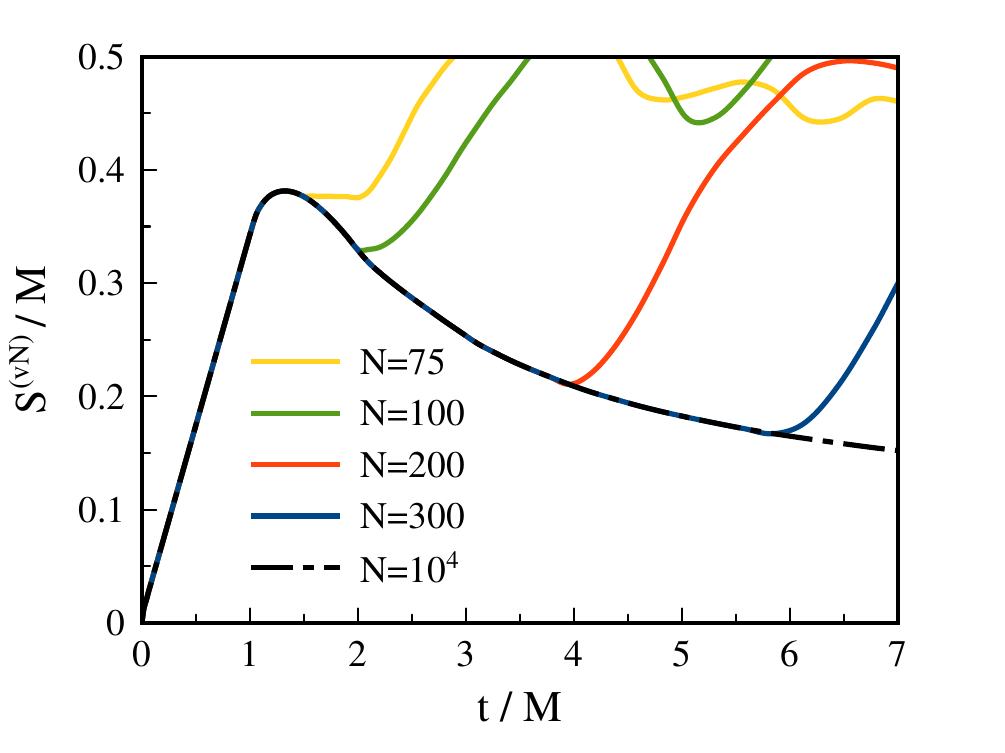}
\caption{\label{fig_SvN_finite_size}
Von Neumann entanglement entropy as a function of time for varying environment sizes~$N$. The system size is $M=50$, other parameters 
$t_{\rm env}=t_{\rm sys}=1, g=0.65$. For this set of parameters the decrease of entropy after the Page time starts to set in for $N/M\approx 1.5$ and is clearly noticeable once 
$N/M\gtrsim 4$. The eventual increase of the entanglement entropy for even longer times (as the curves start to deviate from the $N=10^4$ curve) is due to the system not emptying out completely into the environment and particles being reflected at the right boundary of the environment.
}
\end{figure}

{\it Interpretation and generalizations.} 
From the point of view of an observer in the system~$\cal S$ the dynamical behavior of the entanglement or the energy variance is not surprising. Initially, system plus environment are in an unentangled product state (\ref{eq_initialstate}). The process of emitting particles into the environment generates a complicated state in the system~$\cal S$ that is entangled with the environment, but ultimately this process continues for so long that the system~$\cal S$ is driven into a very small effective Hilbert space ${\cal H}^{\rm (eff)}_{\rm sys}(t\rightarrow\infty)$. Therefore entanglement and environment energy variance are forced to decrease again since the asymptotic small effective Hilbert space does not permit larger values. Fig.~\ref{fig_effH} gives a schematic picture of this process. Notice that in our model we have 
${\rm dim}\,{\cal H}^{\rm (eff)}_{\rm sys}(t\rightarrow\infty)=1$ for $N\gg M^2$ and hence $S^{\rm (vN)}(t\rightarrow\infty)=0$. However, even for smaller ratios $N/M^2$ one can expect to see Page curve behavior in the sense that the entanglement entropy decreases after the Page time, albeit it will not decrease all the way to zero if ${\rm dim}\,{\cal H}^{\rm (eff)}_{\rm sys}(t\rightarrow\infty)>1$. Such finite size behavior is depicted in Fig.~\ref{fig_SvN_finite_size} and would be important for experimental realizations where one might not achieve $N/M^2\rightarrow 0$. Likewise the observations made here will carry over to interacting systems or higher dimensions since the mechanism in Fig.~\ref{fig_effH} is universal (however, then one is numerically limited to much smaller systems).

While the decrease of the entanglement entropy after the Page time $t_{\rm P}$ can easily be understood from the system point of view, a non-omniscient observer in the environment will find it puzzling: The environment keeps absorbing particles that have the curious property of reducing its entanglement entropy and decreasing the energy variance. This behavior is reminiscent of gas particles carefully reassembling themselves in one half of a gas cylinder that they were initially released from. Of course one can always achieve a decrease of entropy in time-reversal invariant systems by first running time backwards to $-t_0<0$ from an initially ordered state $|\Psi(0)\rangle$, and then using this state $|\Psi(-t_0) \rangle$ as the starting point for forward time evolution: In the time interval $[-t_0,0]$ the entropy then has to decrease. However, this decrease as a function of time is very unstable for generic systems and even small perturbations will drive the entropy up again in 
the time interval $[-t_0,0]$. This apparent contradiction between the second law of thermodynamics and microscopic time-reversal invariance was at the center of the Boltzmann-Loschmidt debate, and has been resolved by the understanding of chaotic behavior in generic many-particle systems. Another possibility to generate a decreasing entanglement entropy exists in integrable models with (nearly) linear dispersion relation: For appropriately chosen finite environment sizes one can achieve a refocusing of the quasiparticles that leads to periodic dips of the entanglement entropy \cite{Surace_2020}.

In contrast to such fine-tuned and highly sensitive scenarios with a decreasing entropy, models that follow Fig.~\ref{fig_effH} like the exactly solvable model analyzed in this paper show a robust decrease of the entanglement entropy after the Page time. Notice that at the Page time the entanglement entropy can be made arbitrarily large by increasing~$M$. Even time-dependent perturbations to the system or the environment will not change the qualitative behavior of decreasing entanglement entropy after the Page time as long as the system~$\cal S$ empties into the environment, i.e.~as long as Fig.~\ref{fig_effH} remains applicable. An observer limited to the environment with no knowledge of the full initial state would conclude 
from such a robust decrease of the entanglement entropy that one is watching a movie running backwards. The dynamical buildup of long range entanglement between ${\cal H}_{\rm sys}$ and ${\cal H}_{\rm env}$ up to the Page time is ultimately responsible for the robust Page curve behavior: Particles emitted into the environment after the Page time carry entanglement across the boundary leading to the bending down of the Page curve. This behavior is robust as long as time evolution is unitary and one does not couple to additional external degrees of freedom. It should be emphasized that one cannot make similarly general statements about a bending down of coarse-grained entropies based on Fig.~\ref{fig_effH} since this will depend on the specific coarse-graining procedure employed \cite{RevModPhys.93.035002}.

Interestingly, in our model the bending down of the von Neumann and all other Renyi entanglement entropies can be related to a quantum phase transition of the entanglement Hamiltonian accompanied by non-analytic behavior of the min-entropy, see (\ref{eq_2ndderiv}) and Fig.~\ref{fig_Smin_transition}. Therefore the increasing and decreasing part of the Page curve are separated by a critical point of $S^{\rm (min)}$ (strictly speaking, the critical point slightly precedes the Page time, so this is a qualitative statement about the overall behavior). 
Currently it is unclear whether this observation of a dynamical transition can be generalized to the other forcing mechanisms from 
Fig.~\ref{fig_effH}, which would be very intriguing. At least for models with a conserved charge this seems very likely since the quantum phase transition of the entanglement Hamiltonian is driven by a level crossing between different sectors of the conserved charge, which is unaffected by interactions. More work will be required to explore these questions.  

{\it Summary.}
This paper introduced an exactly solvable model as an example for the general scenario depicted in Fig.~\ref{fig_effH} resulting in a Page curve. Unitary time evolution leads to an increase of the entanglement entropy until the Page time consistent with the semiclassical propotionality between particle current and entanglement generation (\ref{eq_dSdt}). At the Page time the entanglement entropy is proportional to the number of particles~$M$ initially in the system~${\cal S}$. Beyond the Page time the entanglement entropy decreases again and the semiclassical picture breaks down. 
The scenario depicted in Fig.~\ref{fig_effH} generates entanglement dynamics different from the saturation at the volume law scenario commonly discussed in the literature without the need for any fine tuning. 

We have therefore constructed a solvable model that illustrates Polchinski's burning piece of coal \cite{Polchinski_2016}: The early photons are entangled with the remaining coal, but when the coal has burned completely the outgoing photons must be in a pure state (assuming the coal was initially in a pure state).  While this is fundamentally different from a black hole due to the lack of an event horizon \cite{Polchinski_2016}, one can make two observations in this exactly solvable model that have analogues in black hole physics: i)~The semiclassical picture (\ref{eq_dSdt}) breaks down for a subtle quantity like the entanglement entropy once the system~$\cal S$ in ${\cal H}_{\rm sys}$ has evolved to a sufficiently complex state at the Page time \cite{Akers_2022}. ii)~The $q\rightarrow\infty$ Renyi entropy $S^{\rm (min)}$ exhibits non-analytic behavior at a critical time which slightly precedes the Page time (Fig.~\ref{fig_Smin_transition}). The sudden onset of a non-analytic contribution according
to (\ref{eq_Smin_ktau}) shows a certain resemblance to the island formula and the resulting non-analytic behavior in holographic models \cite{Almheiri_2019,Pennington_2022}. In our model the 
non-analytic contribution can be traced back to the sudden emergence of a new closest (in the sense of the Fubini-Study metric) separable state (\ref{eq_closest}) to the time evolved initial state. Whether there is an underlying connection between these observations in holography and our model is a question for future study. In any case the quantum phase transition of the entanglement Hamiltonian discussed above is driven by a new mechanism and deserves further study from the quantum many-body point of view, especially regarding 
generalizations to interacting models with or without conserved charges and the implications for the finite-$q$ Renyi entropies.

Finally, one related interesting property of the state in the environment is that its energy variance decreases after the Page time (Fig.~\ref{fig_DeltaHenv_filling}), corresponding to fewer eigenstates of $H_{\rm env}$ contributing to it. Asymptotically for $t\gg t_{\rm P}$ the energy variance 
$\Delta H_{\rm env}^2$ becomes independent of the number of particles~$M$ in the environment (\ref{eq_low_variance}), which is different from the usual linear dependence on~$M$ for independently injected particles into ${\cal H}_{\rm env}$. Such states with reduced energy uncertainty at nonzero excitation energy could be of interest experimentally.

{\it Acknowledgments.}
I thank  S.~Gopalakrishnan, D.~Huse, R.~Jha, S.~Sachdev and L.~Tagliacozzo for valuable discussions. I am particularly grateful to B.~Halperin whose questions led to the investigation of possible non-analytic behavior. 
This work is funded by
the Deutsche Forschungsgemeinschaft (DFG, German Research Foundation)-217133147/SFB 1073 (Project No. B07). 
It was performed in part at the Aspen Center for Physics, which is supported by National Science Foundation grant PHY-2210452, 
and at the Kavli Institute for Theoretical Physics (KITP), supported by grants NSF PHY-1748958 and PHY-2309135, whose hospitality is gratefully acknowledged. 

\bibliography{Page_curve.bib}

\begin{thebibliography}{37}%
\makeatletter
\providecommand \@ifxundefined [1]{%
 \@ifx{#1\undefined}
}%
\providecommand \@ifnum [1]{%
 \ifnum #1\expandafter \@firstoftwo
 \else \expandafter \@secondoftwo
 \fi
}%
\providecommand \@ifx [1]{%
 \ifx #1\expandafter \@firstoftwo
 \else \expandafter \@secondoftwo
 \fi
}%
\providecommand \natexlab [1]{#1}%
\providecommand \enquote  [1]{``#1''}%
\providecommand \bibnamefont  [1]{#1}%
\providecommand \bibfnamefont [1]{#1}%
\providecommand \citenamefont [1]{#1}%
\providecommand \href@noop [0]{\@secondoftwo}%
\providecommand \href [0]{\begingroup \@sanitize@url \@href}%
\providecommand \@href[1]{\@@startlink{#1}\@@href}%
\providecommand \@@href[1]{\endgroup#1\@@endlink}%
\providecommand \@sanitize@url [0]{\catcode `\\12\catcode `\$12\catcode
  `\&12\catcode `\#12\catcode `\^12\catcode `\_12\catcode `\%12\relax}%
\providecommand \@@startlink[1]{}%
\providecommand \@@endlink[0]{}%
\providecommand \url  [0]{\begingroup\@sanitize@url \@url }%
\providecommand \@url [1]{\endgroup\@href {#1}{\urlprefix }}%
\providecommand \urlprefix  [0]{URL }%
\providecommand \Eprint [0]{\href }%
\providecommand \doibase [0]{http://dx.doi.org/}%
\providecommand \selectlanguage [0]{\@gobble}%
\providecommand \bibinfo  [0]{\@secondoftwo}%
\providecommand \bibfield  [0]{\@secondoftwo}%
\providecommand \translation [1]{[#1]}%
\providecommand \BibitemOpen [0]{}%
\providecommand \bibitemStop [0]{}%
\providecommand \bibitemNoStop [0]{.\EOS\space}%
\providecommand \EOS [0]{\spacefactor3000\relax}%
\providecommand \BibitemShut  [1]{\csname bibitem#1\endcsname}%
\let\auto@bib@innerbib\@empty
\bibitem [{\citenamefont {Eisert}\ \emph {et~al.}(2010)\citenamefont {Eisert},
  \citenamefont {Cramer},\ and\ \citenamefont {Plenio}}]{RevModPhys.82.277}%
  \BibitemOpen
  \bibfield  {author} {\bibinfo {author} {\bibfnamefont {J.}~\bibnamefont
  {Eisert}}, \bibinfo {author} {\bibfnamefont {M.}~\bibnamefont {Cramer}}, \
  and\ \bibinfo {author} {\bibfnamefont {M.~B.}\ \bibnamefont {Plenio}},\
  }\bibfield  {title} {\enquote {\bibinfo {title} {Colloquium: Area laws for
  the entanglement entropy},}\ }\href {\doibase 10.1103/RevModPhys.82.277}
  {\bibfield  {journal} {\bibinfo  {journal} {Rev. Mod. Phys.}\ }\textbf
  {\bibinfo {volume} {82}},\ \bibinfo {pages} {277--306} (\bibinfo {year}
  {2010})}\BibitemShut {NoStop}%
\bibitem [{\citenamefont {Schollw{\"o}ck}(2011)}]{SCHOLLWOCK201196}%
  \BibitemOpen
  \bibfield  {author} {\bibinfo {author} {\bibfnamefont {Ulrich}\ \bibnamefont
  {Schollw{\"o}ck}},\ }\bibfield  {title} {\enquote {\bibinfo {title} {The
  density-matrix renormalization group in the age of matrix product states},}\
  }\href {\doibase https://doi.org/10.1016/j.aop.2010.09.012} {\bibfield
  {journal} {\bibinfo  {journal} {Annals of Physics}\ }\textbf {\bibinfo
  {volume} {326}},\ \bibinfo {pages} {96--192} (\bibinfo {year} {2011})},\
  \bibinfo {note} {january 2011 Special Issue}\BibitemShut {NoStop}%
\bibitem [{\citenamefont {Ryu}\ and\ \citenamefont
  {Takayanagi}(2006)}]{PhysRevLett.96.181602}%
  \BibitemOpen
  \bibfield  {author} {\bibinfo {author} {\bibfnamefont {Shinsei}\ \bibnamefont
  {Ryu}}\ and\ \bibinfo {author} {\bibfnamefont {Tadashi}\ \bibnamefont
  {Takayanagi}},\ }\bibfield  {title} {\enquote {\bibinfo {title} {Holographic
  derivation of entanglement entropy from the anti--de sitter space/conformal
  field theory correspondence},}\ }\href {\doibase
  10.1103/PhysRevLett.96.181602} {\bibfield  {journal} {\bibinfo  {journal}
  {Phys. Rev. Lett.}\ }\textbf {\bibinfo {volume} {96}},\ \bibinfo {pages}
  {181602} (\bibinfo {year} {2006})}\BibitemShut {NoStop}%
\bibitem [{\citenamefont {Calabrese}\ and\ \citenamefont
  {Cardy}(2005)}]{Calabrese_2005}%
  \BibitemOpen
  \bibfield  {author} {\bibinfo {author} {\bibfnamefont {Pasquale}\
  \bibnamefont {Calabrese}}\ and\ \bibinfo {author} {\bibfnamefont {John}\
  \bibnamefont {Cardy}},\ }\bibfield  {title} {\enquote {\bibinfo {title}
  {Evolution of entanglement entropy in one-dimensional systems},}\ }\href
  {\doibase 10.1088/1742-5468/2005/04/P04010} {\bibfield  {journal} {\bibinfo
  {journal} {Journal of Statistical Mechanics: Theory and Experiment}\ }\textbf
  {\bibinfo {volume} {2005}},\ \bibinfo {pages} {P04010} (\bibinfo {year}
  {2005})}\BibitemShut {NoStop}%
\bibitem [{\citenamefont {Ho}\ and\ \citenamefont
  {Abanin}(2017)}]{PhysRevB.95.094302}%
  \BibitemOpen
  \bibfield  {author} {\bibinfo {author} {\bibfnamefont {Wen~Wei}\ \bibnamefont
  {Ho}}\ and\ \bibinfo {author} {\bibfnamefont {Dmitry~A.}\ \bibnamefont
  {Abanin}},\ }\bibfield  {title} {\enquote {\bibinfo {title} {Entanglement
  dynamics in quantum many-body systems},}\ }\href {\doibase
  10.1103/PhysRevB.95.094302} {\bibfield  {journal} {\bibinfo  {journal} {Phys.
  Rev. B}\ }\textbf {\bibinfo {volume} {95}},\ \bibinfo {pages} {094302}
  (\bibinfo {year} {2017})}\BibitemShut {NoStop}%
\bibitem [{\citenamefont {Kim}\ and\ \citenamefont
  {Huse}(2013)}]{PhysRevLett.111.127205}%
  \BibitemOpen
  \bibfield  {author} {\bibinfo {author} {\bibfnamefont {Hyungwon}\
  \bibnamefont {Kim}}\ and\ \bibinfo {author} {\bibfnamefont {David~A.}\
  \bibnamefont {Huse}},\ }\bibfield  {title} {\enquote {\bibinfo {title}
  {Ballistic spreading of entanglement in a diffusive nonintegrable system},}\
  }\href {\doibase 10.1103/PhysRevLett.111.127205} {\bibfield  {journal}
  {\bibinfo  {journal} {Phys. Rev. Lett.}\ }\textbf {\bibinfo {volume} {111}},\
  \bibinfo {pages} {127205} (\bibinfo {year} {2013})}\BibitemShut {NoStop}%
\bibitem [{\citenamefont {Hawking}(1975)}]{Hawking_1975}%
  \BibitemOpen
  \bibfield  {author} {\bibinfo {author} {\bibfnamefont {S.~W.}\ \bibnamefont
  {Hawking}},\ }\bibfield  {title} {\enquote {\bibinfo {title} {Particle
  creation by black holes},}\ }\href@noop {} {\bibfield  {journal} {\bibinfo
  {journal} {Communications in Mathematical Physics}\ }\textbf {\bibinfo
  {volume} {43}},\ \bibinfo {pages} {199} (\bibinfo {year} {1975})}\BibitemShut
  {NoStop}%
\bibitem [{\citenamefont {Page}(1993)}]{PhysRevLett.71.3743}%
  \BibitemOpen
  \bibfield  {author} {\bibinfo {author} {\bibfnamefont {Don~N.}\ \bibnamefont
  {Page}},\ }\bibfield  {title} {\enquote {\bibinfo {title} {Information in
  black hole radiation},}\ }\href {\doibase 10.1103/PhysRevLett.71.3743}
  {\bibfield  {journal} {\bibinfo  {journal} {Phys. Rev. Lett.}\ }\textbf
  {\bibinfo {volume} {71}},\ \bibinfo {pages} {3743--3746} (\bibinfo {year}
  {1993})}\BibitemShut {NoStop}%
\bibitem [{\citenamefont {Page}(2013)}]{Page_2013}%
  \BibitemOpen
  \bibfield  {author} {\bibinfo {author} {\bibfnamefont {Don~N.}\ \bibnamefont
  {Page}},\ }\bibfield  {title} {\enquote {\bibinfo {title} {Time dependence of
  hawking radiation entropy},}\ }\href {\doibase 10.1088/1475-7516/2013/09/028}
  {\bibfield  {journal} {\bibinfo  {journal} {Journal of Cosmology and
  Astroparticle Physics}\ }\textbf {\bibinfo {volume} {2013}},\ \bibinfo
  {pages} {028} (\bibinfo {year} {2013})}\BibitemShut {NoStop}%
\bibitem [{\citenamefont {Mathur}(2009)}]{Mathur_2009}%
  \BibitemOpen
  \bibfield  {author} {\bibinfo {author} {\bibfnamefont {Samir~D.}\
  \bibnamefont {Mathur}},\ }\bibfield  {title} {\enquote {\bibinfo {title} {The
  information paradox: a pedagogical introduction},}\ }\href {\doibase
  10.1088/0264-9381/26/22/224001} {\bibfield  {journal} {\bibinfo  {journal}
  {Classical and Quantum Gravity}\ }\textbf {\bibinfo {volume} {26}},\ \bibinfo
  {pages} {224001} (\bibinfo {year} {2009})}\BibitemShut {NoStop}%
\bibitem [{\citenamefont {Almheiri}\ \emph {et~al.}(2021)\citenamefont
  {Almheiri}, \citenamefont {Hartman}, \citenamefont {Maldacena}, \citenamefont
  {Shaghoulian},\ and\ \citenamefont {Tajdini}}]{RevModPhys.93.035002}%
  \BibitemOpen
  \bibfield  {author} {\bibinfo {author} {\bibfnamefont {Ahmed}\ \bibnamefont
  {Almheiri}}, \bibinfo {author} {\bibfnamefont {Thomas}\ \bibnamefont
  {Hartman}}, \bibinfo {author} {\bibfnamefont {Juan}\ \bibnamefont
  {Maldacena}}, \bibinfo {author} {\bibfnamefont {Edgar}\ \bibnamefont
  {Shaghoulian}}, \ and\ \bibinfo {author} {\bibfnamefont {Amirhossein}\
  \bibnamefont {Tajdini}},\ }\bibfield  {title} {\enquote {\bibinfo {title}
  {The entropy of hawking radiation},}\ }\href {\doibase
  10.1103/RevModPhys.93.035002} {\bibfield  {journal} {\bibinfo  {journal}
  {Rev. Mod. Phys.}\ }\textbf {\bibinfo {volume} {93}},\ \bibinfo {pages}
  {035002} (\bibinfo {year} {2021})}\BibitemShut {NoStop}%
\bibitem [{\citenamefont {Hawking}(1976)}]{PhysRevD.14.2460}%
  \BibitemOpen
  \bibfield  {author} {\bibinfo {author} {\bibfnamefont {S.~W.}\ \bibnamefont
  {Hawking}},\ }\bibfield  {title} {\enquote {\bibinfo {title} {Breakdown of
  predictability in gravitational collapse},}\ }\href {\doibase
  10.1103/PhysRevD.14.2460} {\bibfield  {journal} {\bibinfo  {journal} {Phys.
  Rev. D}\ }\textbf {\bibinfo {volume} {14}},\ \bibinfo {pages} {2460--2473}
  (\bibinfo {year} {1976})}\BibitemShut {NoStop}%
\bibitem [{\citenamefont {Papadodimas}\ and\ \citenamefont
  {Raju}(2013)}]{Papadodimas_2013}%
  \BibitemOpen
  \bibfield  {author} {\bibinfo {author} {\bibfnamefont {K.}~\bibnamefont
  {Papadodimas}}\ and\ \bibinfo {author} {\bibfnamefont {S.}~\bibnamefont
  {Raju}},\ }\bibfield  {title} {\enquote {\bibinfo {title} {An infalling
  observer in ads/cft},}\ }\href@noop {} {\bibfield  {journal} {\bibinfo
  {journal} {J. High Energ. Phys.}\ }\textbf {\bibinfo {volume} {2013}},\
  \bibinfo {pages} {212} (\bibinfo {year} {2013})}\BibitemShut {NoStop}%
\bibitem [{\citenamefont {Lunin}\ and\ \citenamefont
  {Mathur}(2002)}]{LUNIN2002342}%
  \BibitemOpen
  \bibfield  {author} {\bibinfo {author} {\bibfnamefont {Oleg}\ \bibnamefont
  {Lunin}}\ and\ \bibinfo {author} {\bibfnamefont {Samir~D.}\ \bibnamefont
  {Mathur}},\ }\bibfield  {title} {\enquote {\bibinfo {title} {Ads/cft duality
  and the black hole information paradox},}\ }\href {\doibase
  https://doi.org/10.1016/S0550-3213(01)00620-4} {\bibfield  {journal}
  {\bibinfo  {journal} {Nuclear Physics B}\ }\textbf {\bibinfo {volume}
  {623}},\ \bibinfo {pages} {342--394} (\bibinfo {year} {2002})}\BibitemShut
  {NoStop}%
\bibitem [{\citenamefont {Almheiri}\ \emph {et~al.}(2013)\citenamefont
  {Almheiri}, \citenamefont {Marolf}, \citenamefont {Polchinski},\ and\
  \citenamefont {Scully}}]{Almheiri_2013}%
  \BibitemOpen
  \bibfield  {author} {\bibinfo {author} {\bibfnamefont {A.}~\bibnamefont
  {Almheiri}}, \bibinfo {author} {\bibfnamefont {D.}~\bibnamefont {Marolf}},
  \bibinfo {author} {\bibfnamefont {J.}~\bibnamefont {Polchinski}}, \ and\
  \bibinfo {author} {\bibfnamefont {J.}~\bibnamefont {Scully}},\ }\bibfield
  {title} {\enquote {\bibinfo {title} {Black holes: complementarity or
  firewalls?}}\ }\href@noop {} {\bibfield  {journal} {\bibinfo  {journal} {J.
  High Energ. Phys.}\ }\textbf {\bibinfo {volume} {2013}},\ \bibinfo {pages}
  {62} (\bibinfo {year} {2013})}\BibitemShut {NoStop}%
\bibitem [{\citenamefont {Almheiri}\ \emph {et~al.}(2019)\citenamefont
  {Almheiri}, \citenamefont {Engelhardt}, \citenamefont {Marolf},\ and\
  \citenamefont {Maxfield}}]{Almheiri_2019}%
  \BibitemOpen
  \bibfield  {author} {\bibinfo {author} {\bibfnamefont {A.}~\bibnamefont
  {Almheiri}}, \bibinfo {author} {\bibfnamefont {N.}~\bibnamefont
  {Engelhardt}}, \bibinfo {author} {\bibfnamefont {D.}~\bibnamefont {Marolf}},
  \ and\ \bibinfo {author} {\bibfnamefont {H.}~\bibnamefont {Maxfield}},\
  }\bibfield  {title} {\enquote {\bibinfo {title} {The entropy of bulk quantum
  fields and the entanglement wedge of an evaporating black hole},}\
  }\href@noop {} {\bibfield  {journal} {\bibinfo  {journal} {J. High Energ.
  Phys.}\ }\textbf {\bibinfo {volume} {2019}},\ \bibinfo {pages} {63} (\bibinfo
  {year} {2019})}\BibitemShut {NoStop}%
\bibitem [{\citenamefont {Penington}\ \emph {et~al.}(2022)\citenamefont
  {Penington}, \citenamefont {Shenker}, \citenamefont {Stanford},\ and\
  \citenamefont {Yang}}]{Pennington_2022}%
  \BibitemOpen
  \bibfield  {author} {\bibinfo {author} {\bibfnamefont {G.}~\bibnamefont
  {Penington}}, \bibinfo {author} {\bibfnamefont {S.H.}\ \bibnamefont
  {Shenker}}, \bibinfo {author} {\bibfnamefont {D.}~\bibnamefont {Stanford}}, \
  and\ \bibinfo {author} {\bibfnamefont {Z.}~\bibnamefont {Yang}},\ }\bibfield
  {title} {\enquote {\bibinfo {title} {Replica wormholes and the black hole
  interior},}\ }\href@noop {} {\bibfield  {journal} {\bibinfo  {journal} {J.
  High Energ. Phys.}\ }\textbf {\bibinfo {volume} {2022}},\ \bibinfo {pages}
  {205} (\bibinfo {year} {2022})}\BibitemShut {NoStop}%
\bibitem [{\citenamefont {Akers}\ \emph {et~al.}()\citenamefont {Akers},
  \citenamefont {Engelhardt}, \citenamefont {Harlow}, \citenamefont
  {Penington},\ and\ \citenamefont {Vardhan}}]{Akers_2022}%
  \BibitemOpen
  \bibfield  {author} {\bibinfo {author} {\bibfnamefont {C.}~\bibnamefont
  {Akers}}, \bibinfo {author} {\bibfnamefont {N.}~\bibnamefont {Engelhardt}},
  \bibinfo {author} {\bibfnamefont {D.}~\bibnamefont {Harlow}}, \bibinfo
  {author} {\bibfnamefont {G.}~\bibnamefont {Penington}}, \ and\ \bibinfo
  {author} {\bibfnamefont {S.}~\bibnamefont {Vardhan}},\ }\bibfield  {title}
  {\enquote {\bibinfo {title} {The black hole interior from non-isometric codes
  and complexity},}\ }\href {https://doi.org/10.48550/arXiv.2207.06536}
  {\bibinfo  {journal} {Preprint arXiv:2207.06536}\ }\BibitemShut {NoStop}%
\bibitem [{\citenamefont {Liu}\ and\ \citenamefont {Vardhan}(2021)}]{Liu_2021}%
  \BibitemOpen
\bibfield  {journal} {  }\bibfield  {author} {\bibinfo {author} {\bibfnamefont
  {Hong}\ \bibnamefont {Liu}}\ and\ \bibinfo {author} {\bibfnamefont {Shreya}\
  \bibnamefont {Vardhan}},\ }\bibfield  {title} {\enquote {\bibinfo {title} {{A
  dynamical mechanism for the Page curve from quantum chaos}},}\ }\href
  {\doibase 10.1007/jhep03(2021)088} {\bibfield  {journal} {\bibinfo  {journal}
  {Journal of High Energy Physics}\ }\textbf {\bibinfo {volume} {2021}},\
  \bibinfo {pages} {88} (\bibinfo {year} {2021})}\BibitemShut {NoStop}%
\bibitem [{\citenamefont {Lau}\ \emph {et~al.}(2022)\citenamefont {Lau},
  \citenamefont {Noumi}, \citenamefont {Takii},\ and\ \citenamefont
  {Tamaoka}}]{Lau_2022}%
  \BibitemOpen
  \bibfield  {author} {\bibinfo {author} {\bibfnamefont {Pak Hang~Chris}\
  \bibnamefont {Lau}}, \bibinfo {author} {\bibfnamefont {Toshifumi}\
  \bibnamefont {Noumi}}, \bibinfo {author} {\bibfnamefont {Yuhei}\ \bibnamefont
  {Takii}}, \ and\ \bibinfo {author} {\bibfnamefont {Kotaro}\ \bibnamefont
  {Tamaoka}},\ }\bibfield  {title} {\enquote {\bibinfo {title} {{Page curve and
  symmetries}},}\ }\href {\doibase 10.1007/jhep10(2022)015} {\bibfield
  {journal} {\bibinfo  {journal} {Journal of High Energy Physics}\ }\textbf
  {\bibinfo {volume} {2022}},\ \bibinfo {pages} {15} (\bibinfo {year}
  {2022})}\BibitemShut {NoStop}%
\bibitem [{\citenamefont {Chen}\ \emph
  {et~al.}(2020{\natexlab{a}})\citenamefont {Chen}, \citenamefont {Qi},\ and\
  \citenamefont {Zhang}}]{Chen_2020}%
  \BibitemOpen
  \bibfield  {author} {\bibinfo {author} {\bibfnamefont {Yiming}\ \bibnamefont
  {Chen}}, \bibinfo {author} {\bibfnamefont {Xiao-Liang}\ \bibnamefont {Qi}}, \
  and\ \bibinfo {author} {\bibfnamefont {Pengfei}\ \bibnamefont {Zhang}},\
  }\bibfield  {title} {\enquote {\bibinfo {title} {{Replica wormhole and
  information retrieval in the SYK model coupled to Majorana chains}},}\ }\href
  {\doibase 10.1007/jhep06(2020)121} {\bibfield  {journal} {\bibinfo  {journal}
  {Journal of High Energy Physics}\ }\textbf {\bibinfo {volume} {2020}},\
  \bibinfo {pages} {121} (\bibinfo {year} {2020}{\natexlab{a}})}\BibitemShut
  {NoStop}%
\bibitem [{\citenamefont {Chen}\ \emph
  {et~al.}(2020{\natexlab{b}})\citenamefont {Chen}, \citenamefont {Myers},
  \citenamefont {Neuenfeld}, \citenamefont {Reyesc}, ,\ and\ \citenamefont
  {Sandora}}]{Chen_2020b}%
  \BibitemOpen
  \bibfield  {author} {\bibinfo {author} {\bibfnamefont {H.~Z.}\ \bibnamefont
  {Chen}}, \bibinfo {author} {\bibfnamefont {R.~C.}\ \bibnamefont {Myers}},
  \bibinfo {author} {\bibfnamefont {D.}~\bibnamefont {Neuenfeld}}, \bibinfo
  {author} {\bibfnamefont {I.~A.}\ \bibnamefont {Reyesc}}, , \ and\ \bibinfo
  {author} {\bibfnamefont {J.}~\bibnamefont {Sandora}},\ }\bibfield  {title}
  {\enquote {\bibinfo {title} {Quantum extremal islands made easy. part ii.
  black holes on the brane},}\ }\href {https://doi.org/10.1007/JHEP12(2020)025}
  {\bibfield  {journal} {\bibinfo  {journal} {J. High Energ. Phys.}\ ,\
  \bibinfo {pages} {25}} (\bibinfo {year} {2020}{\natexlab{b}})}\BibitemShut
  {NoStop}%
\bibitem [{\citenamefont {Geng}\ \emph {et~al.}(2021)\citenamefont {Geng},
  \citenamefont {L{\"u}st}, \citenamefont {Mishra},\ and\ \citenamefont
  {Wakeham}}]{Geng_2021}%
  \BibitemOpen
  \bibfield  {author} {\bibinfo {author} {\bibfnamefont {H.}~\bibnamefont
  {Geng}}, \bibinfo {author} {\bibfnamefont {S.}~\bibnamefont {L{\"u}st}},
  \bibinfo {author} {\bibfnamefont {R.~K.}\ \bibnamefont {Mishra}}, \ and\
  \bibinfo {author} {\bibfnamefont {D.}~\bibnamefont {Wakeham}},\ }\bibfield
  {title} {\enquote {\bibinfo {title} {Holographic bcfts and communicating
  black holes},}\ }\href {https://doi.org/10.1007/JHEP08(2021)003} {\bibfield
  {journal} {\bibinfo  {journal} {J. High Energ. Phys.}\ }\textbf {\bibinfo
  {volume} {2021}},\ \bibinfo {pages} {3} (\bibinfo {year} {2021})}\BibitemShut
  {NoStop}%
\bibitem [{\citenamefont {Geng}\ \emph {et~al.}(2022)\citenamefont {Geng},
  \citenamefont {Randall},\ and\ \citenamefont {Swanson}}]{Geng_2022}%
  \BibitemOpen
  \bibfield  {author} {\bibinfo {author} {\bibfnamefont {H.}~\bibnamefont
  {Geng}}, \bibinfo {author} {\bibfnamefont {L.}~\bibnamefont {Randall}}, \
  and\ \bibinfo {author} {\bibfnamefont {E.}~\bibnamefont {Swanson}},\
  }\bibfield  {title} {\enquote {\bibinfo {title} {Bcft in a black hole
  background: an analytical holographic model},}\ }\href
  {https://doi.org/10.1007/JHEP12(2022)056} {\bibfield  {journal} {\bibinfo
  {journal} {J. High Energ. Phys.}\ ,\ \bibinfo {pages} {56}} (\bibinfo {year}
  {2022})}\BibitemShut {NoStop}%
\bibitem [{\citenamefont {Heidrich-Meisner}\ \emph {et~al.}(2009)\citenamefont
  {Heidrich-Meisner}, \citenamefont {Manmana}, \citenamefont {Rigol},
  \citenamefont {Muramatsu}, \citenamefont {Feiguin},\ and\ \citenamefont
  {Dagotto}}]{Fabian_2009}%
  \BibitemOpen
  \bibfield  {author} {\bibinfo {author} {\bibfnamefont {F.}~\bibnamefont
  {Heidrich-Meisner}}, \bibinfo {author} {\bibfnamefont {S.~R.}\ \bibnamefont
  {Manmana}}, \bibinfo {author} {\bibfnamefont {M.}~\bibnamefont {Rigol}},
  \bibinfo {author} {\bibfnamefont {A.}~\bibnamefont {Muramatsu}}, \bibinfo
  {author} {\bibfnamefont {A.~E.}\ \bibnamefont {Feiguin}}, \ and\ \bibinfo
  {author} {\bibfnamefont {E.}~\bibnamefont {Dagotto}},\ }\bibfield  {title}
  {\enquote {\bibinfo {title} {Quantum distillation: Dynamical generation of
  low-entropy states of strongly correlated fermions in an optical lattice},}\
  }\href {\doibase 10.1103/PhysRevA.80.041603} {\bibfield  {journal} {\bibinfo
  {journal} {Phys. Rev. A}\ }\textbf {\bibinfo {volume} {80}},\ \bibinfo
  {pages} {041603} (\bibinfo {year} {2009})}\BibitemShut {NoStop}%
\bibitem [{\citenamefont {Peschel}\ and\ \citenamefont
  {Eisler}(2009)}]{Peschel_2009}%
  \BibitemOpen
  \bibfield  {author} {\bibinfo {author} {\bibfnamefont {Ingo}\ \bibnamefont
  {Peschel}}\ and\ \bibinfo {author} {\bibfnamefont {Viktor}\ \bibnamefont
  {Eisler}},\ }\bibfield  {title} {\enquote {\bibinfo {title} {Reduced density
  matrices and entanglement entropy in free lattice models},}\ }\href {\doibase
  10.1088/1751-8113/42/50/504003} {\bibfield  {journal} {\bibinfo  {journal}
  {Journal of Physics A: Mathematical and Theoretical}\ }\textbf {\bibinfo
  {volume} {42}},\ \bibinfo {pages} {504003} (\bibinfo {year}
  {2009})}\BibitemShut {NoStop}%
\bibitem [{\citenamefont {Eisler}\ \emph {et~al.}(2008)\citenamefont {Eisler},
  \citenamefont {Karevski}, \citenamefont {Platini},\ and\ \citenamefont
  {Peschel}}]{Eisler_2008}%
  \BibitemOpen
  \bibfield  {author} {\bibinfo {author} {\bibfnamefont {V}~\bibnamefont
  {Eisler}}, \bibinfo {author} {\bibfnamefont {D}~\bibnamefont {Karevski}},
  \bibinfo {author} {\bibfnamefont {T}~\bibnamefont {Platini}}, \ and\ \bibinfo
  {author} {\bibfnamefont {I}~\bibnamefont {Peschel}},\ }\bibfield  {title}
  {\enquote {\bibinfo {title} {{Entanglement evolution after connecting finite
  to infinite quantum chains}},}\ }\href {\doibase
  10.1088/1742-5468/2008/01/p01023} {\bibfield  {journal} {\bibinfo  {journal}
  {Journal of Statistical Mechanics: Theory and Experiment}\ }\textbf {\bibinfo
  {volume} {2008}},\ \bibinfo {pages} {P01023} (\bibinfo {year}
  {2008})}\BibitemShut {NoStop}%
\bibitem [{\citenamefont {Eisler}\ \emph {et~al.}(2009)\citenamefont {Eisler},
  \citenamefont {Igl{\'o}i},\ and\ \citenamefont {Peschel}}]{Eisler_2009b}%
  \BibitemOpen
  \bibfield  {author} {\bibinfo {author} {\bibfnamefont {Viktor}\ \bibnamefont
  {Eisler}}, \bibinfo {author} {\bibfnamefont {Ferenc}\ \bibnamefont
  {Igl{\'o}i}}, \ and\ \bibinfo {author} {\bibfnamefont {Ingo}\ \bibnamefont
  {Peschel}},\ }\bibfield  {title} {\enquote {\bibinfo {title} {Entanglement in
  spin chains with gradients},}\ }\href {\doibase
  10.1088/1742-5468/2009/02/P02011} {\bibfield  {journal} {\bibinfo  {journal}
  {Journal of Statistical Mechanics: Theory and Experiment}\ }\textbf {\bibinfo
  {volume} {2009}},\ \bibinfo {pages} {P02011} (\bibinfo {year}
  {2009})}\BibitemShut {NoStop}%
\bibitem [{\citenamefont {Eisler}\ and\ \citenamefont
  {Peschel}(2012)}]{Eisler_2012}%
  \BibitemOpen
  \bibfield  {author} {\bibinfo {author} {\bibfnamefont {Viktor}\ \bibnamefont
  {Eisler}}\ and\ \bibinfo {author} {\bibfnamefont {Ingo}\ \bibnamefont
  {Peschel}},\ }\bibfield  {title} {\enquote {\bibinfo {title} {{On
  entanglement evolution across defects in critical chains}},}\ }\href
  {\doibase 10.1209/0295-5075/99/20001} {\bibfield  {journal} {\bibinfo
  {journal} {Europhysics Letters}\ }\textbf {\bibinfo {volume} {99}},\ \bibinfo
  {pages} {20001} (\bibinfo {year} {2012})}\BibitemShut {NoStop}%
\bibitem [{\citenamefont {Capizzi}\ and\ \citenamefont
  {Eisler}(2023)}]{Capizzi_2023}%
  \BibitemOpen
  \bibfield  {author} {\bibinfo {author} {\bibfnamefont {Luca}\ \bibnamefont
  {Capizzi}}\ and\ \bibinfo {author} {\bibfnamefont {Viktor}\ \bibnamefont
  {Eisler}},\ }\bibfield  {title} {\enquote {\bibinfo {title} {{Entanglement
  evolution after a global quench across a conformal defect}},}\ }\href
  {\doibase 10.21468/SciPostPhys.14.4.070} {\bibfield  {journal} {\bibinfo
  {journal} {SciPost Phys.}\ }\textbf {\bibinfo {volume} {14}},\ \bibinfo
  {pages} {070} (\bibinfo {year} {2023})}\BibitemShut {NoStop}%
\bibitem [{\citenamefont {Guinea}\ \emph {et~al.}(1985)\citenamefont {Guinea},
  \citenamefont {Hakim},\ and\ \citenamefont {Muramatsu}}]{Guinea_1985}%
  \BibitemOpen
  \bibfield  {author} {\bibinfo {author} {\bibfnamefont {F.}~\bibnamefont
  {Guinea}}, \bibinfo {author} {\bibfnamefont {V}~\bibnamefont {Hakim}}, \ and\
  \bibinfo {author} {\bibfnamefont {A}~\bibnamefont {Muramatsu}},\ }\bibfield
  {title} {\enquote {\bibinfo {title} {{Bosonization of a two-level system with
  dissipation}},}\ }\href {\doibase 10.1103/physrevb.32.4410} {\bibfield
  {journal} {\bibinfo  {journal} {Physical Review Letters}\ }\textbf {\bibinfo
  {volume} {32}},\ \bibinfo {pages} {4410 -- 4418} (\bibinfo {year}
  {1985})}\BibitemShut {NoStop}%
\bibitem [{\citenamefont {Langreth}\ and\ \citenamefont
  {Nordlander}(1991)}]{Langreth_1991}%
  \BibitemOpen
  \bibfield  {author} {\bibinfo {author} {\bibfnamefont {David~C.}\
  \bibnamefont {Langreth}}\ and\ \bibinfo {author} {\bibfnamefont
  {P.}~\bibnamefont {Nordlander}},\ }\bibfield  {title} {\enquote {\bibinfo
  {title} {{Derivation of a master equation for charge-transfer processes in
  atom-surface collisions}},}\ }\href {\doibase 10.1103/physrevb.43.2541}
  {\bibfield  {journal} {\bibinfo  {journal} {Physical Review B}\ }\textbf
  {\bibinfo {volume} {43}},\ \bibinfo {pages} {2541--2557} (\bibinfo {year}
  {1991})}\BibitemShut {NoStop}%
\bibitem [{\citenamefont {Brandao}\ and\ \citenamefont
  {Datta}(2011)}]{Brandao_2011}%
  \BibitemOpen
  \bibfield  {author} {\bibinfo {author} {\bibfnamefont {Fernando}\
  \bibnamefont {Brandao}}\ and\ \bibinfo {author} {\bibfnamefont {Nilanjana}\
  \bibnamefont {Datta}},\ }\bibfield  {title} {\enquote {\bibinfo {title}
  {One-shot rates for entanglement manipulation under non-entangling maps},}\
  }\href@noop {} {\bibfield  {journal} {\bibinfo  {journal} {IEEE Trans. Inf.
  Theo.}\ }\textbf {\bibinfo {volume} {57}},\ \bibinfo {pages} {1754} (\bibinfo
  {year} {2011})}\BibitemShut {NoStop}%
\bibitem [{\citenamefont {Lockhart}\ and\ \citenamefont
  {Steiner}(2002)}]{Lockhart2002}%
  \BibitemOpen
  \bibfield  {author} {\bibinfo {author} {\bibfnamefont {R.~B.}\ \bibnamefont
  {Lockhart}}\ and\ \bibinfo {author} {\bibfnamefont {M.~J.}\ \bibnamefont
  {Steiner}},\ }\bibfield  {title} {\enquote {\bibinfo {title} {Preserving
  entanglement under perturbation and sandwiching all separable states},}\
  }\href@noop {} {\bibfield  {journal} {\bibinfo  {journal} {Phys. Rev. A}\
  }\textbf {\bibinfo {volume} {65}},\ \bibinfo {pages} {022107} (\bibinfo
  {year} {2002})}\BibitemShut {NoStop}%
\bibitem [{\citenamefont {Chandran}\ \emph {et~al.}(2014)\citenamefont
  {Chandran}, \citenamefont {Khemani},\ and\ \citenamefont
  {Sondhi}}]{Chandran2014}%
  \BibitemOpen
  \bibfield  {author} {\bibinfo {author} {\bibfnamefont {Anushya}\ \bibnamefont
  {Chandran}}, \bibinfo {author} {\bibfnamefont {Vedika}\ \bibnamefont
  {Khemani}}, \ and\ \bibinfo {author} {\bibfnamefont {S.~L.}\ \bibnamefont
  {Sondhi}},\ }\bibfield  {title} {\enquote {\bibinfo {title} {How universal is
  the entanglement spectrum?}}\ }\href {\doibase
  10.1103/PhysRevLett.113.060501} {\bibfield  {journal} {\bibinfo  {journal}
  {Phys. Rev. Lett.}\ }\textbf {\bibinfo {volume} {113}},\ \bibinfo {pages}
  {060501} (\bibinfo {year} {2014})}\BibitemShut {NoStop}%
\bibitem [{\citenamefont {Surace}\ \emph {et~al.}(2020)\citenamefont {Surace},
  \citenamefont {Tagliacozzo},\ and\ \citenamefont {Tonni}}]{Surace_2020}%
  \BibitemOpen
  \bibfield  {author} {\bibinfo {author} {\bibfnamefont {Jacopo}\ \bibnamefont
  {Surace}}, \bibinfo {author} {\bibfnamefont {Luca}\ \bibnamefont
  {Tagliacozzo}}, \ and\ \bibinfo {author} {\bibfnamefont {Erik}\ \bibnamefont
  {Tonni}},\ }\bibfield  {title} {\enquote {\bibinfo {title} {Operator content
  of entanglement spectra in the transverse field ising chain after global
  quenches},}\ }\href {\doibase 10.1103/PhysRevB.101.241107} {\bibfield
  {journal} {\bibinfo  {journal} {Phys. Rev. B}\ }\textbf {\bibinfo {volume}
  {101}},\ \bibinfo {pages} {241107} (\bibinfo {year} {2020})}\BibitemShut
  {NoStop}%
\bibitem [{\citenamefont {Polchinski}(2016)}]{Polchinski_2016}%
  \BibitemOpen
  \bibfield  {author} {\bibinfo {author} {\bibfnamefont {Joseph}\ \bibnamefont
  {Polchinski}},\ }\bibfield  {title} {\enquote {\bibinfo {title} {The black
  hole information problem},}\ }in\ \href {\doibase 10.1142/9789813149441_0006}
  {\emph {\bibinfo {booktitle} {New Frontiers in Fields and Strings}}}\
  (\bibinfo  {publisher} {World Scientific},\ \bibinfo {year}
  {2016})\BibitemShut {NoStop}%
\end{thebibliography}%

\end{document}